\documentclass[aps, prd, showkeys, twocolumn, superscriptaddress, nofootinbib, floatfix]{revtex4-2}

\usepackage{amsmath}
\usepackage{graphicx}
\usepackage{dcolumn}
\usepackage{bm}

\usepackage[bookmarksnumbered, pdfpagelabels=true, plainpages=false, colorlinks=true, linkcolor=blue, citecolor=blue, urlcolor=blue]{hyperref}

\usepackage{url}
\usepackage{xcolor}
\usepackage[T1]{fontenc}
\usepackage{amssymb}
\usepackage{subcaption}

\begin{document}

\title{Radial Oscillations of Viscous Neutron Stars: Zero Diffusion Case}

\author{Raissa F. P. Mendes}
\email{rfpmendes@id.uff.br}
\affiliation{Instituto de F\'{\i}sica, Universidade Federal Fluminense, Niter\'{o}i, Rio de Janeiro, 24210-346, Brazil}
\affiliation{CBPF -- Centro Brasileiro de Pesquisas F\'{\i}sicas, 22290-180, Rio de Janeiro, RJ, Brazil.}

\author{Amanda Guerrieri}
\email{amguerrieri@cbpf.br}
\affiliation{CBPF -- Centro Brasileiro de Pesquisas F\'{\i}sicas, 22290-180, Rio de Janeiro, RJ, Brazil.}

\author{Jo\~ao V. M. Muniz}
\email{joaomotta@id.uff.br}
\affiliation{Instituto de F\'{\i}sica, Universidade Federal Fluminense, Niter\'{o}i, Rio de Janeiro, 24210-346, Brazil}

\author{Gabriel S. Rocha}
\email{gabrielsr@id.uff.br}
\affiliation{Instituto de F\'{\i}sica, Universidade Federal Fluminense, Niter\'{o}i, Rio de Janeiro, 24210-346, Brazil}
\affiliation{Department of Physics and Astronomy, Vanderbilt University, Nashville, TN 37240, USA}

\author{Gabriel S. Denicol}
\email{gsdenicol@id.uff.br}
\affiliation{Instituto de F\'{\i}sica, Universidade Federal Fluminense, Niter\'{o}i, Rio de Janeiro, 24210-346, Brazil}

\begin{abstract}
The spectrum of radial oscillations of neutron stars is systematically studied within two frameworks of viscous relativistic hydrodynamics: the relativistic Navier–Stokes and Israel–Stewart theories. A correspondence is established between the discrete stellar eigenmodes and the continuous dispersion relation of perturbations around a homogeneous fluid, providing a basis for interpreting our numerical results. We analyze the Newtonian limit and assess the impact of relativistic corrections, such as the gravitational redshifting of microscopic relaxation timescales. We show that bulk viscosity can significantly affect the behavior of both hydrodynamic and nonhydrodynamic fundamental modes, and that, depending on the magnitude of the viscous effects, it is the nonhydrodynamic mode that becomes unstable beyond the turning point in a sequence of equilibrium configurations. These results provide a useful step toward systematic studies of neutron star quasinormal modes in the presence of viscosity.
\end{abstract}

\maketitle

\section{Introduction}

Neutron stars serve as unique astrophysical laboratories for probing the properties of dense nuclear matter and strong gravity effects.
While a perfect-fluid description is generally sufficient to represent their equilibrium structure, dynamical phenomena—such as oscillations or interactions in binary systems—can drive the stellar fluid out of local thermodynamic equilibrium, requiring a non-ideal hydrodynamic treatment.
For instance, both shear and bulk viscosities are crucial in determining the $r$-mode instability window \cite{Chandrasekhar:1970pjp, Friedman:1978hf}, with bulk viscosity dominating at higher temperatures \cite{Sawyer:1989dp,Jones:2001ya,Lindblom:2001hd,Haskell:2015iia}. 

Viscosity may also play a role in the post-merger phase of a binary neutron star merger. These events are rich sources of physical information, as demonstrated by the detection of GW170817 and its electromagnetic counterparts \cite{LIGOScientific:2017vwq,LIGOScientific:2017ync}. With the enhanced kilohertz sensitivity of next-generation gravitational-wave observatories such as the Einstein Telescope \cite{Abac:2025saz}, the oscillatory dynamics of the remnant neutron star should be revealed \cite{Stergioulas:2011gd,Bauswein:2011tp,Takami:2014zpa,Bernuzzi:2015rla,Rezzolla:2016nxn}, offering valuable insights into the nuclear equation of state and, potentially, the transport properties of dense nuclear matter.

The potential influence of viscous effects in a binary neutron star merger is highly dependent on the characteristic timescales involved.
When active on secular timescales, viscosity can lead to the braking of differential rotation, thereby influencing mass ejection and the eventual fate of the remnant \cite{Duez:2004nf, Shibata:2017jyf}. If active on shorter, dynamical timescales (in the millisecond range), viscous effects could alter the gravitational-wave signal emitted during the late inspiral \cite{Ripley:2023qxo,Ripley:2023lsq} and post-merger phases. 
Estimates for the impact of energy diffusion, shear viscosity, and bulk viscosity on the post-merger gravitational-wave emission suggest that, under specific conditions, both neutrino-induced shear viscosity and bulk viscous damping arising from deviations from weak chemical equilibrium may leave imprints on the signal \cite{Alford:2017rxf,Most:2021zvc}. As a consequence, viscous effects are increasingly being incorporated into fully nonlinear numerical relativity simulations \cite{Most:2022yhe, Camelio:2022fds, Camelio:2022ljs, Chabanov:2023abq}.

An important tool to describe the oscillation dynamics of neutron stars, and to characterize the gravitational-wave emission during the ringdown phase of a binary neutron star coalescence, is provided by linear perturbation theory. 
The study of quasi-normal modes of neutron stars constitutes a substantial body of work (see, e.g., the review papers \cite{Kokkotas:1999bd,Nollert:1999ji}); however, most of it is restricted to perfect-fluid models or treats viscous effects in a phenomenological manner. A challenge in consistently extending perturbative studies to the non-ideal case lies in the absence of a unique, universally accepted theory of viscous relativistic hydrodynamics, with several possible, competing frameworks in use \cite{Rocha:2023ilf}. In particular, radial perturbations within the relativistic Navier-Stokes theory were studied in Ref.~\cite{Barta:2019tpv}, while the full set of perturbation equations governing non-radial oscillations within the Bemfica–Disconzi–Noronha–Kovtun (BDNK) theory \cite{Bemfica:2017wps, Bemfica:2019knx, Bemfica:2020zjp, Bemfica:2022dnk, Kovtun:2019hdm} was recently derived in Ref.~\cite{Redondo-Yuste:2024vdb}.

In the present work, we investigate the influence of shear and bulk viscosity on the radial oscillation spectrum of neutron stars using linear perturbation theory. This analysis is motivated by several factors. First, linear radial perturbations are a key tool in assessing the stability of neutron stars \cite{Chandrasekhar:1964zz,Chandrasekhar:1964zza}, a subject that has recently been extended to incorporate viscous effects \cite{Caballero:2025omv}. Second, while radial oscillations do not emit gravitational waves directly, it is well established that the quasi-linear coupling between the fundamental radial mode and the fundamental quadrupolar ($l=2$) mode can leave a detectable imprint on the post-merger gravitational-wave signal of binary neutron star collisions \cite{Stergioulas:2011gd, Takami:2014zpa, Rezzolla:2016nxn}. In this context, linear mode analyses may serve as a point of comparison for fully nonlinear simulations, which have already been performed for viscous fluids in spherical symmetry \cite{Camelio:2022fds, Chabanov:2023abq}. Finally, radial oscillations offer the simplest relativistic setting in which the full mode structure—including both hydrodynamic and nonhydrodynamic (dissipative) modes—can be systematically explored. The primary aim of this work is to conduct such an analysis.

For this purpose, we consider two theories of relativistic dissipative hydrodynamics: the relativistic Navier–Stokes \cite{Eckart:1940te,landau:59fluid} and Israel–Stewart \cite{Israel:1976tn,Israel:1979wp} formalisms, which provide descriptions of dissipation of first- and second-order in a gradient expansion, respectively. The background configuration is assumed to be a spherically symmetric, static star, modeled as a perfect fluid with a cold equation of state. The system is then linearly perturbed, with dissipative currents—bulk viscosity $\Pi$, shear viscosity $\pi^{\mu\nu}$, and energy diffusion $Q^\mu$—appearing as perturbative variables. Focusing on the zero-diffusion case, the system of equations simplifies considerably. The resulting linearized equations are then numerically integrated in both the frequency and time domains, from which the discrete set of mode frequencies $\{\omega_{(n)}\}$ can be computed, with $n \in \mathbb{N}$ denoting the overtone number.

The numerical determination of $\{\omega_{(n)}\}$ requires specifying the transport coefficients associated with the adopted relativistic hydrodynamics framework---namely, the bulk and shear viscosity coefficients $\zeta$ and $\eta$, and, in the case of Israel–Stewart theory, the additional relaxation timescales $\tau_\Pi$ and $\tau_\pi$. Several studies have addressed the first-principle computation of these transport coefficients, e.g.~\cite{Haensel:1992zz,Shternin:2008es,Yang:2023ogo,Gavassino:2023xkt}. Here, we instead adopt simpler, yet physically reasonable, prescriptions for the transport coefficients, which in particular allow us to clarify the role of certain singular points that arise in the frequency-domain form of our perturbation equations. Specifically, the bulk and shear viscosity coefficients are assumed to be proportional to the thermodynamical pressure, $\eta = \tau_\eta \textrm{p}$ and $\zeta = \tau_\zeta \textrm{p}$, and two prescriptions are considered for the timescales $\tau_i$ ($i\in \{\eta, \zeta, \Pi, \pi\}$): one in which they are held constant and another in which the redshifted timescales, $t_i = \tau_i e^{-\Phi/2}$, are fixed instead.

To aid the interpretation of our results, we proceed in a step-by-step, pedagogical manner, beginning with preliminary analyses of perturbations around an infinite, homogeneous fluid and around a Newtonian star. In the case of Navier-Stokes theory, we find that the radial mode spectrum of Newtonian and mildly relativistic stars is well approximated by
\begin{equation} \label{eq:omega_homogeneous_ansatz}
    \omega_{(n)} = \omega^\textrm{PF}_{(n)} \left[ -i \frac{n}{n_\textrm{visc}} \pm \sqrt{1 - \left( \frac{n}{n_\textrm{visc}} \right)^2} \right],
\end{equation}
a quantized analogue of the Navier-Stokes dispersion relation for longitudinal perturbations around a homogeneous background \cite{Denicol:2021}\footnote{This structure closely parallels the complex frequencies of a damped harmonic oscillator, $$ \ddot{x} + 2\beta \dot{x} +\omega_0 x = 0,$$ which are given by $\omega_\pm = \omega_0 \left[ -ia \pm \sqrt{1 - a^2} \right]$, with the damping ratio $a \equiv \beta/\omega_0$. }. Here, $\omega_{(n)}^\textrm{PF}$ denotes the perfect-fluid eigenfrequencies and $n_\textrm{visc} \propto \tau_{\eta,\zeta}^{-1}$. For $n < n_\textrm{visc}$ the modes are underdamped and oscillatory, whereas for $n > n_\textrm{visc}$ they are overdamped and purely decaying.

Within Israel–Stewart theory, additional families of modes (including nonhydrodynamic branches) appear since shear and bulk viscosity constitute independent dynamical degrees of freedom. In the shear-viscous case, we find that the radial modes of relativistic stars are approximately given by the roots of 
\begin{equation} \label{eq:IS_ansatz}
    [\omega_{(n)}^2-\omega^2_{\textrm{PF}(n)}](1-i\omega_{(n)}\tau_\pi) + 2i\omega_{(n)} \,\omega^\textrm{PF}_{(n)} \left(\frac{n}{n_\textrm{visc}}\right) = 0,
\end{equation}
which is the quantized analogue of the Israel-Stewart dispersion relation for longitudinal perturbations around a homogeneous state \cite{Denicol:2021}. 
Shear viscosity has little effect on the frequencies of the fundamental ($n=0$) hydrodynamic and nonhydrodynamic modes, with the latter being well described by the infinite-wavelength ($k=0$) limit of the homogeneous-fluid dispersion relation. This contrasts with the general expectation that the finite stellar radius should constrain the maximum allowed wavelength of perturbations.

The situation is markedly different for bulk viscosity. Although the homogeneous-fluid dispersion relation is formally identical in the shear-only and bulk-only sectors, we find that, for both Newtonian and relativistic stars with bulk viscosity, the validity of Eq.~\eqref{eq:IS_ansatz} is typically restricted to high overtone numbers. Low overtones (particularly the fundamental modes) are instead strongly affected by finite-size effects. In addition, the fundamental mode frequency in the bulk-viscous case is significantly modified by relativistic corrections. Interestingly, we find that, depending on the magnitude of bulk-viscous effects, it is the nonhydrodynamic mode that first becomes unstable at the threshold of linear instability associated with gravitational collapse.
Other relativistic effects – such as the gravitational redshifting of microscopic relaxation timescales – are also discussed.

This work is organized as follows. In Sec.~\ref{sec:hydro-models}, we outline the relativistic fluid models considered in this study: the ideal fluid, the relativistic Navier-Stokes theory, and the Israel-Stewart theory. Section~\ref{sec:sph_symmetry} introduces spherical symmetry in the nonlinear gravity-plus-fluid system, while Sec.~\ref{sec:linear_perturbations} specializes to linear radial perturbations around a spherically symmetric, static background. 
Our main results are presented in Sec.~\ref{sec:results}, with a concluding discussion in Sec.~\ref{sec:conclusion}. We employ natural units so that $G = c = k_{B} = 1$ unless stated otherwise.

\section{Relativistic hydrodynamic models}
\label{sec:hydro-models}

In the hydrodynamic regime, where microscopic and macroscopic degrees of freedom possess widely separated characteristic length scales, a given system can be described in terms of macroscopic conserved currents. In neutron star models, the dynamics is usually provided by local conservation laws for baryon net-charge, energy and momentum, dynamically coupled to the spacetime geometry as determined by the Einstein field equations, i.e.,
\begin{subequations}
 \label{eq:basic-hydro-EoM-1}
\begin{align}
\label{eq:hydro-EoM-n-1}
\nabla_\mu N^\mu &= 0, 
\\
\label{eq:hydro-EoM-eps-1}
\nabla_\mu T^{\mu\nu} & = 0,
\\
\label{eq:EE}
G_{\mu \nu} \equiv R_{\mu \nu} - \frac{1}{2} g_{\mu \nu} R & = 8 \pi T_{\mu \nu},
\end{align}
\end{subequations}
where $N^\mu$ is the baryon four-current and $T^{\mu\nu}$ is the energy-momentum tensor. In general, both of these quantities can be decomposed in terms of a normalized timelike four-vector $u^{\mu}$ as
\begin{subequations}
\label{eq:decomps-N-T}
\begin{align}
\label{eq:Ngeneral}
    N^\mu &= \mathcal{N} u^\mu + J^\mu,\\
\label{eq:Tmunugeneral}
    T^{\mu\nu} &= \mathcal{E} u^\mu u^\nu + \Delta^{\mu\nu} \mathcal{P} + 2 Q^{(\mu} u^{\nu)} + \pi^{\mu\nu},
\end{align}    
\end{subequations} 
with parentheses denoting symmetrization over the indices.
The vector $u^\mu$ will be later identified with the fluid four-velocity once thermodynamic-frame conditions are imposed. 
In the above expressions, $\Delta^{\mu \nu} \equiv g^{\mu\nu} + u^\mu u^\nu$ denotes the projection tensor onto the subspace orthogonal to $u^\mu$, and by definition $u_\mu J^\mu = u_\mu Q^\mu = u_\mu \pi^{\mu\nu} = 0$, with the (symmetric) tensor $\pi^{\mu\nu}$ further defined as traceless, $\pi^\mu_{\,\,\mu} = 0$.
The individual components of the conserved currents can be obtained as
\begin{equation}
\label{eq:from-T-to-comps}
\begin{aligned}
\mathcal{N} & = - u_{\mu} N^{\mu}, \quad 
\mathcal{E} = T^{\mu \nu} u_{\mu} u_{\nu},
\quad
\mathcal{P} = \frac{1}{3} \Delta_{\mu \nu} T^{\mu \nu}
\\
Q^{\mu} & = - \Delta^{\mu}_{\ \nu} T^{\nu \lambda} u_{\lambda}, \quad 
J^{\mu} = \Delta^{\mu}_{\ \nu} N^{\nu}, \quad
\pi^{\mu \nu} = \Delta^{\mu \nu \alpha \beta} T_{\alpha \beta},
\end{aligned}    
\end{equation}
where $\Delta_{\alpha\beta}^{\mu\nu} \equiv \Delta_{(\alpha}^\mu \Delta_{\beta)}^\nu - \frac{1}{3} \Delta^{\mu\nu} \Delta_{\alpha\beta}$ is the doubly-symmetric and traceless projector. The quantities defined in Eq.~\eqref{eq:from-T-to-comps} represent, respectively, the total net particle density, the total energy density, the total isotropic pressure, the energy diffusion current, the particle diffusion current, and the shear-stress tensor or anisotropic pressure tensor, all measured in the local rest frame of the fluid (once a thermodynamic frame condition is specified; see below).

If the system is in local equilibrium, we have $\mathcal{N} = \mathrm{n}$, $\mathcal{E} = \mathrm{e}$, $\mathcal{P} = \mathrm{p}$, $J^{\mu} = 0$, $Q^{\mu} = 0$, $\pi^{\mu \nu} = 0$, since in this state there should be no diffusion currents and pressure should be spatially isotropic in the local rest frame. The functional form of the thermodynamic pressure $\mathrm{p} = \mathrm{p}(\mathrm{n},\mathrm{e})$ is specified by the underlying equation of state, which contains information on the relevant microscopic interactions of the system. Then the equations of motion read 
\begin{subequations}
 \label{eq:euler-EoM}
\begin{align}
\label{eq:euler-EoM-n}
u^{\mu} \nabla_{\mu}\mathrm{n} + \mathrm{n} \theta & = 0,
\\
\label{eq:euler-EoM-eps}
u^\mu \nabla_\mu \mathrm{e} + (\mathrm{e}+ \mathrm{p}) \theta &= 0, \\
\label{eq:euler-EoM-umu}
(\mathrm{e} + \mathrm{p})u^{\lambda}\nabla_{\lambda}u^{\mu} + \Delta^{\mu \nu}\nabla_{\nu}\mathrm{p}  &= 0,
\end{align}
\end{subequations}
where $\theta \equiv \nabla_\mu u^\mu$ is the expansion rate. The above equations, together with Eq.~\eqref{eq:EE}, form a closed system of partial differential equations. This is in general no longer true if dissipation is present, as dissipative currents introduce additional degrees of freedom.

In general configurations, conserved currents can be decomposed into equilibrium and non-equilibrium (dissipative) parts. This separation is a theoretical artifact for the description of the system, whose physical information is contained in the total $N^{\mu}$ and $T^{\mu \nu}$. In practice, such a decomposition requires a prescription to define the reference local equilibrium state. In this work, we shall adopt the prescription by C.~Eckart \cite{Eckart:1940te}, according to which the total particle number density and energy density are such that they satisfy the equilibrium equation of state, i.e., $\mathcal{N} \equiv \mathrm{n}$ and $\mathcal{E} \equiv \mathrm{e}$. This choice defines what is meant by temperature, $T$, and chemical potential $\mu$ in non-equilibrium configurations. Moreover, in Eckart's frame, the fluid four-velocity is defined such that a co-moving observer detects no particle diffusion, implying $J^{\mu} \equiv 0$. Another popular prescription, commonly employed in applications for heavy-ion collisions, is that of Landau \cite{landau:59fluid}, according to which a co-moving observer should detect no energy diffusion, implying $Q^{\mu} \equiv 0$ in that case. For both prescriptions, the total isotropic pressure is $\mathcal{P} \equiv \mathrm{p} + \Pi$, where, once again, $\mathrm{p} = \mathrm{p}(\mathrm{n},\mathrm{e})$ and $\Pi$ is the bulk viscous pressure. 

Then, with these assumptions, inserting the general decompositions \eqref{eq:decomps-N-T} into Eqs.~\eqref{eq:hydro-EoM-n-1}, \eqref{eq:hydro-EoM-eps-1} and considering Eckart's frame, we obtain
\begin{widetext}
\begin{subequations}
 \label{eq:basic-hydro-EoM}
\begin{align}
\label{eq:hydro-EoM-n}
u^{\mu} \nabla_{\mu}\mathrm{n} + \mathrm{n} \theta & = 0,
\\
\label{eq:hydro-EoM-eps}
u^\mu \nabla_\mu \mathrm{e} + (\mathrm{e} + \mathrm{p} + \Pi) \theta + \nabla_{\mu} Q^{\mu} 
 + Q^{\nu} u^\mu \nabla_\mu u_{\nu} + \pi^{\mu \nu} \sigma_{\mu\nu}  & =  0, \\
\label{eq:hydro-EoM-umu}
(\mathrm{e} + \mathrm{p} + \Pi)u^{\lambda}\nabla_{\lambda}u^{\mu} + \Delta^{\mu \nu}\nabla_{\nu}(\mathrm{p} + \Pi)  
+ Q^{\mu} \theta + Q^{\lambda} \nabla_{\lambda}u^{\mu} +  \Delta^{\mu \nu} u^\lambda \nabla_\lambda Q_{\nu} + \Delta^{\mu \nu} \nabla_{\lambda}\pi^{\lambda}_{ \ \nu} & = 0 ,
\end{align}
\end{subequations}
\end{widetext}
where we defined the shear tensor, $\sigma^{\mu\nu} \equiv \Delta_{\alpha \beta}^{\mu\nu} \nabla^\alpha u^\beta$. In order to close the system of partial differential equations, one must either introduce new dynamical equations for the dissipative currents, $\Pi$, $Q^{\mu}$, and $\pi^{\mu \nu}$, or specify constitutive relations that connect these currents to the thermodynamic variables $\textrm{n}$, $\textrm{e}$, and $u^{\mu}$. Different prescriptions for doing so define different theories of relativistic dissipative hydrodynamics (see Ref.~\cite{Rocha:2023ilf} for a recent review). In the following subsections, we outline the specific theories to be employed in the present work: relativistic Navier-Stokes and Israel-Stewart.

\subsection{Relativistic Navier-Stokes}

The first theory that we shall explore is the relativistic extension of the Navier-Stokes equations. It can be motivated within a gradient expansion, where dissipative currents at a given time are assumed to arise due to spatial inhomogeneities in the fluid configuration. For instance, an inhomogeneous temperature profile naturally induces a finite heat flux. The relativistic Navier-Stokes equations arise by including all possible first-order spacelike derivatives consistent with the symmetries of the dissipative currents:
\begin{equation}
\label{eq:NS}
\Pi = - \zeta \nabla_\mu u^\mu, 
\quad
Q^\lambda = \lambda_{Q} \Delta_\nu^\lambda \nabla^\nu \left(\frac{\mu}{T}\right),
\quad
\pi^{\mu\nu} = -2\eta \sigma^{\mu\nu},  
\end{equation}
where $\zeta$ is the bulk viscosity, $\lambda_{Q}$ is the heat flux coefficient and $\eta$ is the shear viscosity. These transport coefficients also encode microscopic information about the system and can be derived from power-counting procedures from the underlying non-equilibrium dynamics \cite{deGroot:80relativistic,cercignani:02relativistic,Denicol:2021}. Even though widely employed in non-relativistic applications, relativistic Navier-Stokes equations are known to allow for superluminal propagation of perturbations \cite{pichon:65etude} and to predict instabilities of perturbations around global equilibrium \cite{hiscock:85generic,Hiscock:1983zz}.

\subsection{Israel-Stewart Theory}

One proposal to circumvent the pathologies of the relativistic Navier-Stokes theory was put forward in Refs.~\cite{Israel:1976tn,Israel:1979wp}, by Israel and Stewart. Recognizing that the culprit for the pathologies within Navier-Stokes stems from the instantaneous response of the fluid to spatial gradients, their approach promotes the dissipative currents to independent dynamical variables which follow relaxation-type equations, 
\begin{subequations}
\label{eq:IS}
\begin{align}
\tau_\Pi u^\mu \nabla_\mu \Pi + \Pi & = - \zeta \nabla_\mu u^\mu + ...,
\\
    \tau_Q \Delta_\nu^\lambda u^\mu \nabla_\mu Q^\nu + Q^\lambda & = \lambda_{Q} \Delta_\nu^\lambda \nabla^\nu \left(\frac{\mu}{T}\right) + ...,
\\    
    \tau_\pi \Delta^{\mu \nu}_{\alpha \beta} u^\lambda \nabla_\lambda \pi^{\alpha \beta} + \pi^{\mu\nu} &= -2\eta \sigma^{\mu\nu} + ...,    
\end{align}    
\end{subequations}
where the ellipses represent possible coupling terms between the dissipative currents themselves and/or terms involving derivatives of the thermodynamic variables, which emerge in systematic power counting procedures \cite{Wagner:2022ayd,Fotakis:2022usk}. For simplicity, these terms will be neglected in the present work. We also note that, in this minimal formulation, Navier-Stokes constitutive relations can be readily recovered in the limit $\tau_\Pi, \tau_Q, \tau_\pi \to 0$. 

The restoration of causality and stability in the Israel–Stewart formulation is conditional. For instance, the following constraint on the transport coefficients can be obtained by imposing these requirements \cite{Pu:2009fj}:
\begin{equation}
\label{eq:lin-causality}
\begin{aligned}
&    
c_{s}^{2}
+
\frac{1}{\textrm{e} + \textrm{p}} \left(\frac{4}{3} \frac{\eta}{\tau_{\pi}} +
\frac{\zeta}{\tau_{\Pi}}\right)
\leq
1
.
\end{aligned}    
\end{equation}
Although derived for linear perturbations around global equilibrium and within the Landau frame, the above expression will serve as a guide for selecting physical coefficients in Sec.~\ref{sec:results} below. 

More recently, an alternative approach to circumvent the inconsistencies of the relativistic Navier–Stokes theory was proposed by Bemfica, Disconzi, Noronha, and Kovtun \cite{Bemfica:2017wps, Bemfica:2019knx, Bemfica:2020zjp, Bemfica:2022dnk, Kovtun:2019hdm}. In this framework, the constitutive relations are allowed to include time-like derivatives, such as $u^{\mu} \nabla_{\mu}T$. However, this modification requires alternative matching conditions \cite{Bemfica:2017wps,Salazar:2024tut}, distinct from those in the Eckart or Landau frames, in order to ensure causality and stability. We defer the analysis of such models to future work.

\section{Dissipative hydrodynamics in spherical symmetry}
\label{sec:sph_symmetry}

In this section, we specialize the general hydrodynamic framework described above to systems with spherical symmetry, following the approach of  Refs.~\cite{Gerlach:1979rw,Gundlach:1999bt,Redondo-Yuste:2024vdb}.

\subsection{Spacetime}

We consider a spherically symmetric spacetime, $(\mathcal{M},g_{\mu\nu})$, where $\mathcal{M} = \mathcal{M}^2 \times_r \mathcal{S}^2$ is the warped product of a Lorentzian manifold with metric $g_{AB}$ and a 2-sphere with metric $\gamma_{ab}$. Covering $\mathcal{M}^2$ with coordinates $\{y^A\}$ and $\mathcal{S}^2$ with coordinates $\{\theta^a\}$, the line element can be written as 
\begin{equation}\label{eq:metric}
    ds^2 = g_{AB}(y) dy^A dy^B + r^2(y) \gamma_{ab} d\theta^a d\theta^b,
\end{equation}
where $r(y)$ is the areal radius. Introducing a basis $\{l_A, n_A\}$ on $\mathcal{M}^2$ satisfying $l_A l^A = -1$, $n^A n_A = 1$, and $l^A n_A = 0$, the metric on this submanifold takes the form
\begin{equation}
    g_{AB} = - l_A l_B + n_A n_B.
\end{equation}

The Einstein tensor in a spherically symmetric spacetime can be written as
\begin{equation}
\begin{aligned}
    G_{\mu\nu} dx^\mu dx^\nu & = \frac{1}{2} \left( \mathcal{E}_g g_{AB} + \mathcal{E}_p p_{AB} - \mathcal{E}_q q_{AB} \right) dy^A dy^B \\
    & + r^2 \mathcal{E}_S \gamma_{ab} d\theta^a d\theta^b,
\end{aligned}    
\end{equation}
where $p_{AB} \equiv l_A l_B + n_A n_B$ and $q_{AB} \equiv l_A n_B + n_A l_B$. 
It has the most general form for a symmetric tensor on $\mathcal{M}^2$, where it is expressed in terms of three scalars ($\mathcal{E}_g = g^{AB} G_{AB}$, $\mathcal{E}_p = p^{AB} G_{AB}$, and $\mathcal{E}_q = q^{AB} G_{AB}$). However, as a consequence of spherical symmetry, it is described by a single scalar ($\mathcal{E}_S = \gamma^{ab} G_{ab}/ (2r^2)$) on the sphere. Defining $U \equiv l^A v_A$, $W \equiv n^A v_A$, $\mu \equiv \nabla_A l^A$, and $\nu \equiv \nabla_A n^A$, one obtains
\begin{subequations} \label{eq:Egframe}
\begin{align}
    \mathcal{E}_g & = \frac{2}{r} \left(W' - \dot{U} + \nu W - \mu U + \frac{v^2}{r} - \frac{1}{r} \right), \\
    \mathcal{E}_p & = \frac{2}{r} \left( - W' - \dot{U} + \nu W + \mu U \right), \\
    \mathcal{E}_q & = \frac{2}{r} \left( - \dot{W} - U' + \mu W + \nu U  \right),\\
    \mathcal{E}_S &  = \frac{1}{r} \left( -\dot{U} + W' + \nu W - \mu U -\frac{r}{2} R^{(2)} \right),
\end{align}
\end{subequations}
where we use the notation $f'\equiv \partial_rf$ and $\dot{f} \equiv \partial_t f$ for any scalar quantity $f$.

\subsection{Matter}

Analogously to the Einstein tensor, the energy-momentum tensor can be decomposed as
\begin{equation}
\label{eq:EMT-sph-sym}
\begin{aligned}
    T_{\mu\nu} dx^\mu dx^\nu & = \frac{1}{2} \left( t_g g_{AB} + t_p p_{AB} - t_q q_{AB} \right) dy^A dy^B \\
    &
    + r^2 t_S \gamma_{ab} d\theta^a d\theta^b,
\end{aligned}    
\end{equation}
where spherical symmetry implies, e.g., that there are no anisotropic stresses on the sphere. 

The decomposition above can be connected to the standard form given in Eq.~\eqref{eq:decomps-N-T}. To that end, it is convenient to introduce a local basis on $\mathcal{M}^{2}$ adapted to the fluid, consisting of the fluid four-velocity $u_\mu dx^\mu = u_A dy^A$ and a spacelike vector $m_\mu dx^\mu = m_A dy^A$ such that $m_A m^A = 1$ and $u_A m^A = 0$. 
In the adapted basis, the shear stress tensor and the energy-diffusion 4-current take the form
\begin{align}
    Q_\mu dx^\mu &= Q_A dy^A = Q m_A dy^A, \\
    \pi_{\mu\nu} dx^\mu dx^\nu & = \tilde{\pi} m_A m_B dy^A dy^B - \frac{1}{2} \tilde{\pi} r^2 \gamma_{ab} d\theta^a d\theta^b.
\end{align}
Therefore, the energy-momentum tensor becomes
\begin{equation}
\begin{aligned}
\label{eq:Tmunu_spherical}
    T_{\mu\nu} dx^\mu dx^\nu &= 
    \left[\mathcal{E} u_A u_B + (\mathcal{P} + \tilde{\pi}) m_A m_B 
    \right.
    \\
    &
    \left.
    + 2 Q m_{(A} u_{B)} \right] dy^A dy^B + r^2 \gamma_{ab} (\mathcal{P} -\tilde{\pi}/2)  d\theta^a d\theta^b.
\end{aligned}
\end{equation}
We note that, if the identifications $l_A = u_A$ and $n_A = m_A$ were made, then one would get $t_g = - \mathcal{E} + \mathcal{P} + \tilde{\pi}$, $t_p = \mathcal{E} + \mathcal{P} + \tilde{\pi}$, and $t_q = -2 Q$, but, in general, the correspondence depends on the relation between $\{l^{A},n^{A}\}$ and $\{u^{A},m^{A}\}$. 
Since dissipative currents emerge as a response to inhomogeneities in the fluid, it is also useful to express the gradients of the 4-velocity as
\begin{equation}
\nabla_{\alpha} u_{\beta} dx^\alpha dx^\beta 
=
(\vartheta m_{A} - \kappa u_{A}) m_{B} dy^A dy^B 
+
r X \gamma_{ab}d\theta^a d\theta^b,
\end{equation}
where $\vartheta \equiv \nabla_A u^{A}$, $\kappa \equiv \nabla_A m^{A}$, and $X \equiv u^{A}\nabla_A r$,
which implies that 
\begin{equation}
\begin{aligned}
\theta & = \vartheta + \frac{2 X}{r} ,
\\
\sigma_{\mu \nu} dx^\mu dx^\nu &= 
\frac{2}{3} \left( \vartheta - \frac{X}{r} \right) m_{A} m_{B} dy^A dy^B, 
\\
&
-
\frac{r^{2}}{3} \gamma_{ab} \left( \vartheta - \frac{X}{r} \right) d\theta^a d\theta^b .
\end{aligned}    
\end{equation}

\subsection{Equations of motion in the non-linear regime}

In spherical symmetry, the Einstein field equations read $\mathcal{E}_j = 8 \pi t_j$, with $j \in \{g, p, q, S\}$ and $\mathcal{E}_j$ given by Eq.~\eqref{eq:Egframe}. They are coupled to the local conservation laws \eqref{eq:basic-hydro-EoM}. Using the fluid-adapted basis $\{u^{A},m^{A}\}$, the conservation equations in the spherically symmetric case become
\begin{widetext}
\begin{subequations}
\label{eq:consv_laws_um}
\begin{align}
 u^A \nabla_A \textrm{n} + \textrm{n} \left( \vartheta + \frac{2X}{r}\right) & = 0 ,\label{eq:Ncons}
\\
 u^A \nabla_A \textrm{e} + (\textrm{e} + \mathrm{p} + \Pi) \left( \vartheta + \frac{2X}{r}\right) +  m^A \nabla_A Q + Q \kappa + \frac{2}{r} YQ + \tilde{\pi}\left(\vartheta - \frac{X}{r}\right)&= 0, \label{eq:Econs} \\
(\textrm{e} + \mathrm{p} + \Pi)\kappa
+
m^A \nabla_A(\mathrm{p} + \Pi) 
+ 2 Q \left( \vartheta +  \frac{X}{r}\right)  +  u^A \nabla_A Q  + m^A \nabla_A \tilde{\pi} + \tilde{\pi} \kappa + \frac{3}{r}\tilde{\pi} Y &= 0, \label{eq:pcons}
\end{align}
\end{subequations}
\end{widetext}
where we defined $Y \equiv m^A \nabla_A r$. The above set of equations is complemented either by 
the Israel-Stewart relaxation equations \eqref{eq:IS}, that now read
\begin{equation} \label{eq:IS_spherical}
\begin{aligned}
&
\tau_{\Pi} u^A \nabla_A \Pi + \Pi
=  
- \zeta \left( \vartheta + \frac{2X}{r}\right),
\\
&
\tau_{Q} u^A \nabla_A Q 
+
Q
=
\lambda_{Q} m^A \nabla_A \left( \frac{\mu}{T}\right),
\\
&
\tau_{\pi} u^A \nabla_A \tilde{\pi}
+
\tilde{\pi}
=
-\frac{4}{3} \eta \left( \vartheta - \frac{X}{r} \right),
\end{aligned}    
\end{equation}
or by the Navier-Stokes  constitutive relations for the dissipative currents \eqref{eq:NS}, which can be obtained from Eqs.~\eqref{eq:IS_spherical} by taking $\tau_\Pi, \tau_Q, \tau_\pi \to 0$.

\section{Linear radial perturbations on a static background} \label{sec:linear_perturbations}

In this section we consider radial perturbations around a static, spherically symmetric background.
We adopt Schwarzschild-type coordinates $\{t,r\}$ on $\mathcal{M}^2$ such that
\begin{equation} \label{eq:l_n_in_usua_coords}
    l_A dy^A = - e^{\Phi(t,r)/2} dt, \qquad 
    n_A dy^A = e^{\Lambda(t,r)/2} dr,
\end{equation}
and the metric assumes the form 
\begin{equation}
    g_{AB} dy^A dy^B = - e^{\Phi(t,r)} dt^2 + e^{\Lambda(t,r)} dr^2.
\end{equation}

For the static background, we recover the Tolman-Oppenheimer-Volkoff equations for an ideal fluid; see Appendix~\ref{sec:background}. These determine the background metric potentials $\Phi_{(0)}$ and $\Lambda_{(0)}$, as well as the fluid variables $\textrm{p}_{(0)}$, $\textrm{e}_{(0)}$, and $\textrm{n}_{(0)}$ given a specified equation of state. 

We then consider linear radial perturbations on top of the static background. For the metric potentials, we write 
\begin{align}
    \Phi(t,r) & = \Phi_{(0)}(r) + \delta \Phi (t,r), \\
    \Lambda(t,r) & = \Lambda_{(0)}(r) + \delta \Lambda (t,r).
\end{align}
We construct the energy-momentum tensor as in Eq.~\eqref{eq:Tmunu_spherical}, and the number current density $N^\mu = \mathcal{N} u^\mu$, setting
\begin{align}
    \mathcal{E}(t,r) & = \mathrm{e}(t,r) = \mathrm{e}_{(0)}(r) + \delta \mathrm{e}(t,r), \\
    \mathcal{P}(t,r) &= \textrm{p}_{(0)}(r) + \delta \textrm{p} (t,r) + \Pi (t,r), \\ \mathcal{N}(t,r) &= \mathrm{n}(t,r) = \mathrm{n}_{(0)}(r) + \delta \mathrm{n}(t,r).
\end{align}
The quantities encoding viscosity---$Q(t,r)$, $\tilde{\pi}(t,r)$, and $\Pi(t,r)$---are treated as perturbation variables and kept to first order. 
The perturbation to the four-velocity reads
\begin{equation}
\label{eq:pert-u}
\begin{aligned}
    \delta u_\mu dx^\mu & = \left(\gamma n_A^{(0)} - \frac{1}{2} \delta g_{AB} u^B_{(0)} \right) dy^A \\
    & = e^{\Phi_{(0)}/2} \left( - \frac{1}{2} \delta \Phi dt + e^{\Lambda_{(0)}-\Phi_{(0)}} \dot{\xi} dr \right),
\end{aligned}    
\end{equation}
which ensures that the four-velocity remains properly normalized to first order in the perturbed variables. Here,
\begin{equation}
\gamma \equiv e^{(\Lambda_{(0)} - \Phi_{(0)})/2} \dot{\xi}  
\end{equation}
serves as a definition of the Lagrangian displacement $\xi(t,r)$. Up to linear order, the space-like vector field $m_\mu$ is given by
\begin{equation}
    m_\mu dx^\mu = e^{\Lambda_{(0)}/2} [ -\dot{\xi} dt + (1 + \delta \Lambda/2) dr ].
\end{equation}

We now derive the linearized dynamical equations. From this point onward, we will omit the $(0)$ subscript on background quantities to simplify the notation, as long as there is no risk of ambiguity.
From the Einstein field equations, the $tr$, $rr$, and $tt$ components yield, respectively
\begin{subequations}
\label{eq:EFE-pert}
\begin{align}
\label{eq:dLambdadt}
    \dot{\delta \Lambda} & = - 8\pi r e^{\Lambda} [ (\textrm{e} + \textrm{p}) \dot{\xi} + e^{(\Phi-\Lambda)/2} Q],
\\ 
\delta \Phi' &= 8\pi r e^{\Lambda} (\delta \textrm{p} + \Pi + \tilde{\pi} ) + \frac{e^{\Lambda}}{r}  (1 + 8 \pi r^2 \textrm{p}) \delta \Lambda,
\\
\delta \Lambda' &= 8\pi r e^{\Lambda} \delta \textrm{e} - \delta \Lambda \left( \frac{1}{r} - \Lambda' \right) \notag \\
&
+ \delta \Phi \left( \frac{1-e^{\Lambda}}{r} + 8 \pi r e^{\Lambda} \textrm{e}  - \Lambda' \right).
\end{align}    
\end{subequations}
On the other hand, from the local conservation laws \eqref{eq:consv_laws_um} we can derive the following linearized equations of motion:
\begin{subequations}
\label{eq:consv-pert}
\begin{align}
\label{eq:deltan}
    \frac{\delta \textrm{n}}{\textrm{n}} + \frac{1}{2} \delta \Lambda + \xi \left( \frac{2}{r} + \frac{1}{2} \Lambda' + \frac{\textrm{n}'}{\textrm{n}} \right) + \xi' &= 0,\\
\label{eq:dedt}
    \dot{\delta \textrm{e}} + \frac{1}{2} (\textrm{e} + \textrm{p}) \dot{\delta\Lambda} + \left( \textrm{e}' + \frac{1}{2r} (\textrm{e} + \textrm{p})(4 + r \Lambda') \right) \dot{\xi} 
    \notag
    & \\
    + (\textrm{e} + \textrm{p}) \dot{\xi'} +e^{(\Phi-\Lambda)/2} \left[ Q' + Q \left( \frac{2}{r} + \Phi' \right)\right] &= 0,\\
    e^{\Lambda - \Phi} (\textrm{e} + \textrm{p}) \ddot{\xi} + \frac{1}{2} (\textrm{e} + \textrm{p}) \delta \Phi' + \delta \textrm{p}' + \Pi' + \tilde{\pi}'
     & \notag
    \\
    + \frac{1}{2} \Phi' (\delta \textrm{p} + \delta \textrm{e} + \Pi + \tilde{\pi} )
    + \frac{3}{r} \tilde{\pi}
    + e^{(\Lambda-\Phi)/2} \dot{Q} & = 0, 
\end{align}
\end{subequations} 
where equation \eqref{eq:deltan} is obtained by integrating in $t$, a simplification made possible by the use of the Eckart frame \cite{Misner:1973prb}. Equations~\eqref{eq:EFE-pert} and \eqref{eq:consv-pert} must be supplemented by the constitutive relations provided by the chosen hydrodynamic theory. The equations of motion in the Israel-Stewart formalism, Eq.~\eqref{eq:IS_spherical}, yield 
\begin{subequations}
\label{eq:IS-NS-eoms}
\begin{align}
\tau_\Pi^{(0)} \dot{\Pi} + e^{\Phi/2} \Pi = & - \zeta^{(0)} \left[ \dot{\xi}' + \frac{1}{2}\dot{\delta \Lambda} + \dot{\xi} \left( \frac{2}{r} +\frac{\Lambda'}{2} \right)
    \right], \\
\label{eq:IS-NS-eoms-Q}    
    \tau_Q^{(0)} \dot{Q} + e^{\Phi/2} Q = & e^{(\Phi-\Lambda)/2} \lambda_Q (\mu/T)',
    \\
    \tau_\pi^{(0)} \dot{\tilde{\pi}} + e^{\Phi/2} \tilde{\pi} = & - \frac{4}{3} \eta^{(0)} \left[ \dot{\xi}' + \frac{1}{2}\dot{\delta \Lambda} + \dot{\xi} \left(- \frac{1}{r} +\frac{\Lambda'}{2} \right)
    \right],    
\end{align}    
\end{subequations}
with the Navier-Stokes constitutive relations being recovered in the $\tau_\Pi^{(0)}, \tau_Q^{(0)} , \tau_\pi^{(0)} \to 0$ limit. Equations \eqref{eq:EFE-pert}, \eqref{eq:consv-pert} and \eqref{eq:IS-NS-eoms} form a closed system of linear partial differential equations for the variables $\{ \xi, \delta \Lambda, \delta \Phi, \delta \textrm{n}, \delta \textrm{e}, \Pi, Q, \tilde{\pi} \}$, which are functions of $t$ and $r$, with radial-dependent coefficients $\{\textrm{n}, \textrm{e}, \Phi, \Lambda\}$, related to the background fluid, and $\{\textrm{p}, \zeta^{(0)}, \lambda_Q, \eta^{(0)}, \tau_\Pi^{(0)}, \tau_Q^{(0)} , \tau_\pi^{(0)} \}$, which are functions of $(\textrm{n},\textrm{e})$ (or, equivalently, $(\mu, T)$) related both to the equation of state and transport properties, which encode microscopic information of the system. From the equation of state, we can also relate $\delta \textrm{p}$ to $\delta \textrm{e}$ through $\delta \textrm{p} = c_s^2(r) \delta \textrm{e}$, where $c_s^2$ is the speed of sound. 

Now we turn our attention to certain features of the system \eqref{eq:EFE-pert}-\eqref{eq:IS-NS-eoms}, which motivate further simplifications of the equations of motion. In the absence of energy diffusion, $Q = 0$, the following simplifications apply: $(i)$   Eq.~\eqref{eq:dLambdadt} can be directly integrated, yielding $\delta \Lambda = - 8\pi r e^{\Lambda} (\textrm{e} + \textrm{p}) \xi$, which allows us to eliminate $\delta \Lambda$ from the remaining equations; $(ii)$ Eq.~\eqref{eq:dedt} can also be directly integrated, relating $\delta \textrm{e}$ algebraically to $\delta \Lambda$ and $\xi$. Indeed, it can be directly verified that the resulting equation is equivalent to the ``first law of thermodynamics'' in the form $\delta \textrm{e} = (\textrm{e} + \textrm{p}) \delta \textrm{n} / \textrm{n}$, when Eq.~\eqref{eq:deltan} is applied (as well as $\textrm{e}' = (\textrm{e} + \textrm{p}) \textrm{n}'/\textrm{n}$). 
When $Q \neq 0$, simplifications $(i)$ and $(ii)$ are no longer valid. Besides, the right-hand side of Eq.~\eqref{eq:IS-NS-eoms-Q} is not automatically zero at the background if there are spatial gradients of temperature or chemical potential, and this equation must then be treated with care. 
Therefore, for the sake of simplicity, we shall assume from the next subsection onward that the diffusion coefficient vanishes identically $Q(t,r) = 0$, leaving its treatment to future studies.

\subsection{Zero energy diffusion limit}
\label{eq:zero-ener-diff}

In the zero energy diffusion limit, the equations derived in the last section simplify considerably, and we obtain a master equation for the Lagrangian displacement,
\begin{widetext}
\begin{align} \label{eq:xiQ0}
   0 & = (\textrm{e} + \textrm{p})e^{\Lambda-\Phi} \ddot{\xi} - c_s^2 (\textrm{e} + \textrm{p}) \xi'' + \Pi' + \tilde{\pi}' + \xi' (\textrm{e} + \textrm{p}) e^\Lambda \left[ 4\pi r \textrm{p} + \frac{m}{r^2} + c_s^2\left( \frac{5m}{r^2} - \frac{2}{r} - 4\pi r \textrm{e} \right) - e^{-\Lambda} (c_s^2)' \right] \nonumber \\
    & + \Pi e^\Lambda \left[ \frac{m}{r^2} + 4\pi r (\textrm{e} + 2 \textrm{p}) \right]
    + \tilde{\pi} e^\Lambda \left[ \frac{3}{r} - \frac{5m}{r^2} + 4\pi r (\textrm{e} + 2\textrm{p}) \right]
    + 2 \xi (\textrm{e} + \textrm{p}) e^{2\Lambda} \left\{ -16 \pi^2 r^2 \textrm{p}^2 - \frac{m}{r^3} (1 + 8\pi r^2 \textrm{p}) \right.
    \nonumber \\
    & + \left. \frac{m^2}{r^4} + c_s^2 \left[ \frac{1}{r^2} - 2\pi \textrm{p} (1-8\pi r^2 \textrm{e}) - 2\pi \textrm{e}\left( 1- \frac{4m}{r} \right) + \frac{5m}{r^4} (m-r) \right]
    +e^{-\Lambda} (c_s^2)' \left(\frac{5m}{2r^2} - \frac{1}{r} + 2\pi r \textrm{p} \right)   
    \right\},
\end{align}    
\end{widetext}
coupled to 
\begin{subequations}
\label{eq:IS-eoms-Q0}
\begin{align}
\tau_\Pi \dot{\Pi} + e^{\Phi/2} \Pi = & - \zeta \left[ \dot{\xi}' + \dot{\xi} \left( \frac{5-e^{\Lambda}}{2r} -4\pi r e^\Lambda \textrm{p} \right)
    \right], \\
\tau_\pi \dot{\tilde{\pi}} + e^{\Phi/2} \tilde{\pi} = & - \frac{4}{3} \eta \left[ \dot{\xi}' - \dot{\xi} \left( \frac{1 + e^\Lambda}{2 r} + 4\pi r e^\Lambda \textrm{p} \right)
    \right],    
\end{align}    
\end{subequations}
where $m \equiv (r/2)(1-e^{-\Lambda})$ and we dropped the $(0)$ superscript on the transport coefficients.
Regularity at $r=0$ demands that
\begin{equation}
    \xi(t,0) = 0, \qquad \tilde{\pi}(t,0) = 0.
\end{equation}

Equations \eqref{eq:xiQ0} and \eqref{eq:IS-eoms-Q0} can be numerically integrated once initial conditions for $\{\xi,\Pi,\tilde{\pi}\}$ are specified. An example of such an evolution will be presented in Sec.~\ref{sec:relativistic_results}; however, in order to investigate the full fluid spectrum, it is more convenient to perform a frequency domain analysis, assuming a harmonic time dependence for all perturbation variables, such that
\begin{equation}
\begin{aligned}
   \xi(t,r) &= e^{-i\omega t} \xi(r), 
   \\
    \Pi(t,r) &= e^{-i\omega t} \Pi (r),
    \\
    \tilde{\pi} (t,r) &= e^{-i\omega t} \tilde{\pi}(r),
\end{aligned}
\end{equation}
with $\omega \in \mathbb{C}$. Equation~\eqref{eq:xiQ0} then becomes a master second-order ordinary differential equation for the Lagrangian displacement, with bulk and shear viscosity expressed algebraically in terms of $\xi$ and $\xi'$ via
\begin{align} \label{eq:ISfreq}
    \Pi(r) &= \frac{\omega \zeta}{\omega \tau_\Pi + i e^{\Phi/2}} \left[ -\xi' + \xi \left( 4\pi r e^{\Lambda} \textrm{p} + \frac{e^\Lambda - 5}{2r}
    \right) \right], \nonumber \\
    \tilde{\pi} (r) &= \frac{4\omega \eta}{3 (\omega \tau_\pi + i e^{\Phi/2})} \left[ - \xi' + \xi \left( 4\pi r e^{\Lambda} \textrm{p} + \frac{e^\Lambda + 1}{2r}
    \right) \right].    
\end{align}

Physical solutions for the Lagrangian displacement must be regular throughout the domain, which imposes the boundary condition $\xi(r=0) = 0$ and a Robin-type boundary condition (i.e., $a \xi' (R) + b\xi(R) = 0$ for some $a,b\in \mathbb{C}$) at the stellar surface ($r=R$), the precise form of which depends on the choice of transport coefficients. Since in the frequency domain bulk- and shear-stress perturbations are algebraically related to the Lagrangian displacement and its derivative, $\Pi(R)$ and $\tilde{\pi}(R)$ are completely determined by $\xi(R)$ and $\xi'(R)$; no additional boundary conditions for the viscous variables are needed at the stellar surface.

Since the master equation for $\xi$ in the frequency domain is linear and homogeneous, it admits a rescaling freedom: if $\xi$ is a solution, then so is $c\xi$, for any $c \in \mathbb{C}$. Therefore, the boundary conditions at $r=0$ and $r=R$ cannot be simultaneously satisfied for arbitrary values of $\omega$, but only for a (possibly infinite) discrete set of eigenfrequencies, which we denote by $\omega_{(n)}$. Here, the index $n\in\mathbb{N}$ labels the eigenmodes and corresponds to the number of nodes (zeros) in the radial profile of the eigenfunctions.

We note that the prefactor in Eqs.~\eqref{eq:ISfreq} is invariant under complex conjugation combined with the substitution $\operatorname{Re}(\omega) \to -\operatorname{Re}(\omega)$. As a consequence, when these equations are inserted into the frequency-domain form of Eq.~\eqref{eq:xiQ0}, taking the complex conjugate yields a differential equation for $\xi^*$ identical to that for $\xi$, but with a frequency with opposite real part. Therefore, if $\operatorname{Re}(\omega) \neq 0$, mode frequencies occur in pairs $(\omega, -\omega^*)= (\operatorname{Re}(\omega) + i \operatorname{Im}(\omega), -\operatorname{Re}(\omega) + i \operatorname{Im}(\omega))$, with eigenfunctions related by complex conjugation.

\subsubsection{Analysis of singular points} \label{sec:singular}

The frequency-domain form of Eq.~\eqref{eq:xiQ0} can be schematically written as
\begin{equation}
    a(r) \xi'' + b(r) \xi' + c(r) = 0,
\end{equation}
where $a$, $b$, and $c$ are functions of background quantities. In particular, 
\begin{equation}
    a(r) = -c_s^2 (\textrm{e} + \textrm{p}) - \frac{\omega \zeta}{\omega \tau_\Pi + i e^{\Phi/2}} - \frac{4 \omega \eta}{3(\omega \tau_\pi + i e^{\Phi/2})}, 
\end{equation}
and $b$ can be conveniently expressed as
\begin{equation}
    b(r) = a' - 2 e^{\Lambda} a \left[ \frac{2m}{r^2} - \frac{1}{r} - 2 \pi r (\textrm{e} + \textrm{p}) \right].
\end{equation}

Casting the master equation in the canonical form 
\begin{equation}
    \xi'' + p(r) \xi' + q(r) = 0,
\end{equation}
we see that the $p=b/a$ and $q=c/a$ coefficients may develop simple poles at some radial coordinate $r_* \in (0,R)$ if either of the following conditions holds:
\begin{itemize}
    \item[(i)] $a(r_*) = 0$,
    \item[(ii)] $\omega \tau_\Pi(r_*) =-i e^{\Phi(r_*)/2}$,
    \item[(iii)] $\omega \tau_\pi(r_*) =-i e^{\Phi(r_*)/2}$.
\end{itemize}
Condition (ii) arises due to the presence of terms proportional to $(\omega \tau'_\Pi + i e^{\Phi/2} \Phi')/(\omega \tau_\Pi + i e^{\Phi/2})$ in $p$, with an analogous argument also holding for condition (iii).

For either of the above conditions to be satisfied, the frequency $\omega$ must be purely imaginary with a negative imaginary part, i.e., $\omega = -i|\omega|$. Therefore, the presence of singular points within the integration domain affects exclusively the search for purely damped eigenmodes -- although it can be avoided in particular settings, as we discuss below. 

The presence of poles in the coefficients $p$ and $q$ leads to diverging (unphysical) solutions unless specific Robin-type boundary conditions are satisfied at $r = r_*$, namely
\begin{itemize}
    \item[(i)] $a'(r_*) \xi'(r_*) + c(r_*) \xi(r_*) = 0$,
    \item [(ii)] $ \xi'(r_*) - \xi(r_*) \left( 4\pi r_* e^{\Lambda(r_*)} \textrm{p}(r_*) + \frac{e^{\Lambda(r_*)} - 5}{2r_*}
    \right) = 0$,
    \item[(iii)] $\xi'(r_*) - \xi(r_*) \left( 4\pi r_* e^{\Lambda(r_*)} \textrm{p}(r_*) + \frac{e^{\Lambda(r_*)} + 1}{2r_*}
    \right) = 0$,
\end{itemize}
for the three cases considered above.
Simultaneous compliance with the boundary conditions at $r=0$, $r=R$, and $r=r_*$ is not typically possible, but one can still search for---and, in the cases we studied, find---discontinuous solutions with support on either the interval $[0,r_*]$ or $[r_*,R]$ which vanish in the complementary part of the domain.  Similar ``trapped modes'' arise in entirely different physical systems \cite{Chiang:2025gpa}.

\subsubsection{Newtonian limit} \label{sec:Newtonian_limit}

The Newtonian limit\footnote{The Newtonian limit can be obtained by reinstating factors of $c$ and $G$ in Eq.~(\ref{eq:xiQ0}) and formally taking the limit $c \to \infty$.} of Eq.~(\ref{eq:xiQ0}) reads 
\begin{align} \label{eq:xiQ0newtonian}
   0 =& \rho \, \ddot \xi - \left( c_s^2 \rho \xi' \right)' - \frac{2 c_s^2 \rho}{r} \xi' + \Pi' + \tilde{\pi}' + \frac{3 \tilde{\pi}}{r}  \nonumber \\
   +&  2 \xi \rho \left[ - \frac{m}{r^3} + \frac{c_s^2}{r^2} - \frac{(c_s^2)'}{r} \right],
\end{align}
where $\rho = m_b \textrm{n}$ denotes the rest-mass density, with $m_b$ the baryon (neutron) mass.
In the frequency domain, the Israel-Stewart equations \eqref{eq:ISfreq} take the form
\begin{align} \label{eq:ISfreqNP}
    \Pi(r) &= \frac{\omega \zeta}{\omega \tau_\Pi + i} \left( -\xi' - \frac{2\xi}{r} \right), \\
    \label{eq:ISfreqNp}
    \tilde{\pi} (r) & = \frac{4 \omega \eta}{3(\omega \tau_\pi + i)} \left( - \xi' + \frac{\xi}{r} \right) ,
\end{align}
with $\omega \in \mathbb{C}$, and Eq.~\eqref{eq:xiQ0newtonian} can be cast as a second-order ordinary differential equation for the radial profile of the Lagrangian displacement, $\xi(r)$.

\section{Results}
\label{sec:results}

Our goal in this section is to elucidate the main features of the full spectrum of a radially oscillating relativistic star in the absence of heat fluxes, but with shear and bulk viscosity present. To this end, we first examine two simpler toy systems. In Sec.\ref{sec:homogeneous}, we present the dispersion relations for perturbations around an infinite, homogeneous fluid, with a short derivation of the relevant expressions provided in Appendix~\ref{apn:long-pert-hom-fluid}. In Sec.~\ref{sec:Newtonian_results}, we present results for a polytropic star in Newtonian gravity, which can be mapped remarkably well to the previously obtained dispersion relations. We then return in Sec.~\ref{sec:relativistic_results} to the main problem of the present work, beginning with a minimal extension of the Newtonian scenario---intended to discern genuinely relativistic effects---, and finally considering more realistic choices for both the equation of state and transport coefficients.

\subsection{Perturbations around an infinite, homogeneous fluid} \label{sec:homogeneous}

As we shall see, the functional structure of the various (radial) mode frequencies of a (Newtonian or relativistic) star resembles that of longitudinal perturbations around global equilibrium of an infinite, homogeneous fluid in a flat spacetime. In this subsection, we summarize the results pertaining to the latter system, which shall serve as a reference for the forthcoming subsections. The results are summarily derived in Appendix \ref{apn:long-pert-hom-fluid}. 

In contrast to the main problem we are interested in the remainder of this work, we consider a fluid initially in global equilibrium, such that its state is specified by $\{\mathrm{n}_{(0)}, \mathrm{e}_{(0)}, \mathrm{p}_{(0)}(\mathrm{n}_{(0)}, \mathrm{e}_{(0)}), u^{\mu}_{(0)}\}$. Then, it is linearly perturbed to a non-equilibrium state specified by $\{\mathrm{n}_{(0)} + \bar\delta \mathrm{n}, \mathrm{e}_{(0)} + \bar\delta \mathrm{e} , \mathrm{p}_{(0)}(\mathrm{n}_{(0)}, \mathrm{e}_{(0)}) + \bar\delta \mathrm{p} + \bar\Pi, u^{\mu}_{(0)} + \bar\delta u^{\mu}, \bar\pi^{\alpha \beta}\}$, where the bars and the symbol $\bar\delta$ are used to emphasize that the spacetime metric is fixed, and not perturbed as in the remainder of this work.

In general, the perturbations can be decomposed into transverse and longitudinal components with respect to the unit 3-vector in the direction of the perturbation wavevector, whose components in cartesian coordinates are denoted by $\hat{k}^{i}$. For longitudinal modes, we have that, in the rest frame of the background fluid, $\bar\delta u^{i} \propto \hat{k}^{i}$ and $\bar\delta\pi^{i j} \propto \hat{k}^{i} \hat{k}^{j} - (1/3) \delta^{ij}$. Performing a Fourier analysis of the equations of motion linearized in the $\bar\delta$-perturbations yields the following dispersion relation for Navier-Stokes theory,
\begin{equation} \label{eq:NShomogeneous} 
    \omega(k) = c_s k \left[ - i \frac{k}{k_\textrm{visc}} \pm \sqrt{ 1 - \left(\frac{k}{k_\textrm{visc}}\right)^2}\right],
\end{equation}
which holds in the local rest frame of the background fluid, with
\begin{equation} \label{eq:kvisc}
    k_\textrm{visc} \equiv \frac{2 c_s}{\zeta + 4\eta/3} \frac{\textrm{e} + \textrm{p}}{c^2}.
\end{equation}
Systems with bulk-only and shear-only perturbations can be analyzed by simply setting $k_\textrm{visc}= k_\textrm{visc}|_{\eta=0}$ and $k_\textrm{visc}=k_\textrm{visc}|_{\zeta=0}$, respectively. 

For Israel-Stewart theory, the dispersion relation for longitudinal perturbations in a fluid with shear stresses alone follows from the roots of the polynomial \cite{Pu:2009fj,Denicol:2021}
\begin{equation} \label{eq:homogeneous_IS_shear}
    (\omega^2-c_s^2 k^2)(1-i\omega \tau_\pi) + 2i\omega c_s k \left(\frac{k}{k_\textrm{visc}|_{\zeta=0}}\right) = 0.
\end{equation}
On the other hand, for a fluid with only bulk stress, the corresponding dispersion relation is obtained from the roots of the polynomial \cite{Pu:2009fj}
\begin{equation} \label{eq:homogeneous_IS_bulk}
    (\omega^2-c_s^2 k^2)(1-i\omega \tau_\Pi) + 2i\omega c_s k \left(\frac{k}{k_\textrm{visc}|_{\eta=0}}\right) = 0.
\end{equation}
We note that when both bulk and shear perturbations are present, the dispersion relations in Israel-Stewart theory take the form of higher-order polynomials, giving rise to additional families of modes \cite{Pu:2009fj}.

\subsection{Newtonian analysis} \label{sec:Newtonian_results}

To lay the groundwork for the relativistic analysis, it is instructive to first examine an example within Newtonian gravity.
For this purpose, let us consider a polytropic equation of state,
\begin{equation} \label{eq:polytropic}
    \textrm{p} = K \rho^{\Gamma},
\end{equation}
with polytropic exponent $\Gamma = 2$, and where $\rho = m_b \textrm{n}$ again denotes the rest-mass density. The background solution for this case can be obtained in closed form, as described in Appendix~\ref{sec:background}.

The master equations governing radial perturbations of a Newtonian fluid were summarized in Sec.~\ref{sec:Newtonian_limit}.
To proceed, we must specify the transport coefficients. We will assume constant relaxation times, $\tau_\pi = \textrm{cte}$ and $\tau_\Pi = \textrm{cte}$, and adopt the following parametrization:
\begin{equation}
\label{eq:eta-zeta-param}
  \eta = \tau_\eta \textrm{p}, \qquad \zeta = \tau_\zeta \textrm{p},
\end{equation}
where $\tau_{\eta,\zeta} > 0$ have dimensions of time.
This prescription simplifies the frequency-domain analysis since no singular points arise in the interior of the integration domain, except at the critical frequency $\omega_\textrm{cr}$ such that
\begin{equation} \label{eq:wcr_Newt}
    \Gamma + \frac{4 \omega_\textrm{cr} \tau_\eta}{3(\omega_\textrm{cr} \tau_\pi + i)} + \frac{\omega_\textrm{cr} \tau_\zeta}{\omega_\textrm{cr} \tau_\Pi + i} = 0.
\end{equation}
In what follows, we analyze the cases of shear and bulk viscosity separately, comparing the Newtonian results with the homogeneous case presented in Sec.~\ref{sec:homogeneous}.

\begin{figure*}[th]
    \begin{subfigure}[b]{0.5\textwidth}
        \centering
        \includegraphics[width=\linewidth]{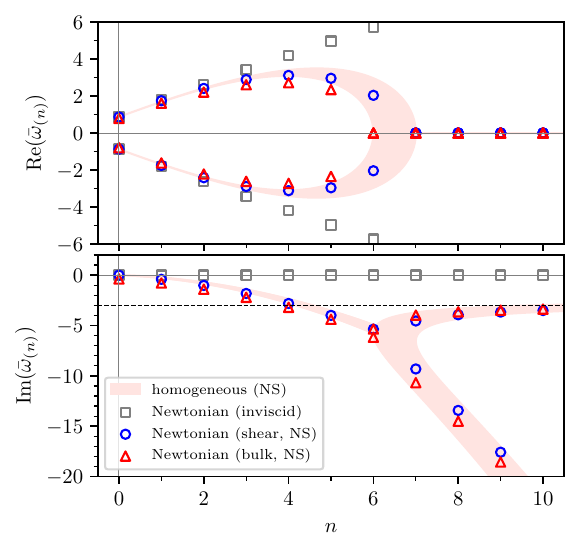}
        \caption{Navier-Stokes (NS)}
        \label{fig:NS_Newtonian_shear_bulk}
    \end{subfigure}%
    ~ 
    \begin{subfigure}[b]{0.5\textwidth}
        \centering
        \includegraphics[width=\linewidth]{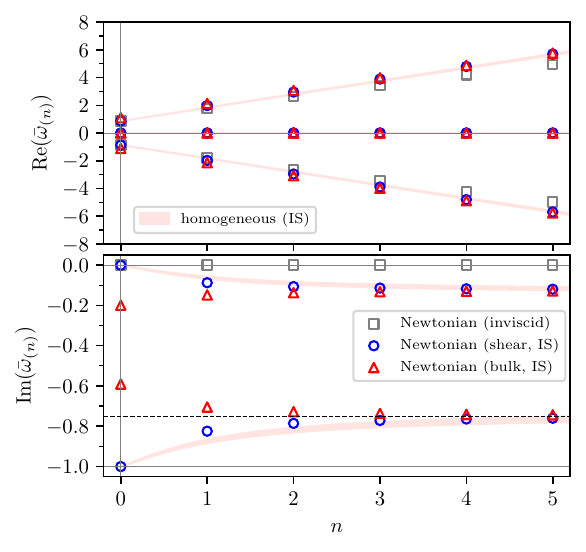}
        \caption{Israel-Stewart (IS)}
        \label{fig:IS_Newtonian_shear_bulk}
    \end{subfigure}
    \caption{Real (upper panels) and imaginary (lower panels) parts of the eigenfrequencies $\bar{\omega}_{(n)} = \omega_{(n)}/\omega_0$ as a function of the overtone number $n$ for a Newtonian $\Gamma = 2$ polytrope subject to shear (blue circles) or bulk (red triangles) viscosity in the Navier-Stokes (left) and Israel-Stewart (right) theories. (a) We consider $\tau_\eta \omega_0 = 0.5$ in the shear viscous case (blue circles) and $\tau_\zeta \omega_0 = 2/3 $ in the bulk viscous case (red triangles), which both correspond to the critical frequency $\omega_{\textrm{cr}} = -3 i \omega_0$ (dashed line). The shaded region corresponds to Eq.~(\ref{eq:omega_homogeneous_ansatz}) with $n_\textrm{cr} \in [6,7]$. (b) We set $\omega_\pi = -i\omega_0$ in the shear viscous case and $\omega_\Pi = -i\omega_0$ in the bulk viscous case, along with the same values of $\tau_{\eta,\zeta}$ as in (a). The critical frequency is now  $\omega_{\textrm{cr}} = -0.75 i \omega_0$ (dashed line). The shaded region corresponds to the roots of Eq.~(\ref{eq:omega_homogeneous_ansatz_IS_S}), with $n_\textrm{cr}$ in the same range as before. In all panels, mode frequencies in the case of zero viscosity (gray squares) are also shown for comparison.}
    \label{fig:Newtonian}
\end{figure*}

\subsubsection{Shear viscosity only}

By inserting the background quantities for a $\Gamma = 2$ polytrope, as described in Appendix~\ref{sec:background}, into Eq.~\eqref{eq:xiQ0newtonian}, the frequency-domain master equation for the Lagrangian displacement takes the following form in the shear viscous case:
\begin{align} \label{eq:newtonian_master_shear}
    0 & =  \xi''+ 2 \cot\bar{r} \xi' + \frac{\xi}{2\bar{r}^2 (\bar{\omega} - \bar{\omega}_{\textrm{cr}}) \bar{\omega}_\pi} \left[ 4 \bar{r} \bar{\omega} (\bar{\omega}_{\textrm{cr}} - \bar{\omega}_\pi) \cot\bar{r} \right.
    \nonumber \\
    & - \left. \bar{\omega}_{\textrm{cr}} (\bar{\omega} - \bar{\omega}_\pi) (4-\bar{r}^3 \bar{\omega}^2 \csc \bar{r}) \right],
\end{align}
where an overbar indicates dimensionless quantities: $\bar{r} = r/r_0$ (with respect to which derivatives are taken), $\bar{\omega} = \omega/\omega_0$, and similarly for the remaining frequencies. The scaling constants are given by $\omega_0 = \sqrt{2\pi \rho_c}$ and $r_0 = \omega_0^{-1} \sqrt{\textrm{p}(\rho_c)/\rho_c}$, with $\rho_c$ denoting the central rest-mass density. Moreover, $\omega_\textrm{cr}$ is given by the root of Eq.~\eqref{eq:wcr_Newt},
\begin{equation} \label{eq:wcr_Newt_eta}
    \omega_{\textrm{cr}} = - \frac{3i \Gamma}{4 \tau_\eta + 3 \Gamma \tau_\pi},
\end{equation}
and we have introduced
\begin{equation} \label{eq:omega_pi}
    \omega_\pi \equiv -\frac{i}{\tau_\pi}.
\end{equation}

For the adopted choice of transport coefficients, Eq.~\eqref{eq:newtonian_master_shear} is regular in the interior of the integration domain, $\bar{r} \in (0,\pi)$, and, at the boundaries, one must impose
\begin{equation} \label{eq:bcN_shear}
\xi(0) = 0, \quad \xi'(\pi) = \frac{\bar{\omega} \xi(\pi) \{4 \bar{\omega}_\pi - \bar{\omega}_{\textrm{cr}} [4 - \pi^2 \bar{\omega} (\bar{\omega} - \bar{\omega}_\pi)] \}}{4\pi \bar{\omega}_\pi (\bar{\omega}-\bar{\omega}_{\textrm{cr}})} ,
\end{equation}
to ensure regularity of physical solutions.

Figure \ref{fig:NS_Newtonian_shear_bulk} shows the first eigenfrequencies within Navier-Stokes theory (i.e., $\tau_\pi = 0$) and for $\omega_{\textrm{cr}} = -3i\omega_0$ as blue circles.
To enable comparison with the analytic result for perturbations around a homogeneous, infinite fluid, we relate the wave-number $k$ in the dispersion relation~\eqref{eq:NShomogeneous} to the overtone number $n$, and adopt the following ansatz, motivated by the structure of  Eq.~\eqref{eq:NShomogeneous}: 
\begin{equation} \label{eq:omega_homogeneous_ansatz}
    \omega_{(n)} = \omega_\textrm{PF}(n) \left[ -i \frac{n}{n_\textrm{visc}} \pm \sqrt{1 - \left( \frac{n}{n_\textrm{visc}} \right)^2} \right],
\end{equation}
where $\omega_\textrm{PF}(n)$ is the Newtonian result for the mode frequencies of a perfect fluid, and $n_\textrm{visc} \propto \tau_\eta^{-1}$ is defined to be the threshold value of $n$ above which modes cease to oscillate and become purely damped, which decreases as $\tau_\eta$ increases. As shown in Fig. \ref{fig:NS_Newtonian_shear_bulk}, the simple ansatz \eqref{eq:omega_homogeneous_ansatz} captures very well the overall behavior of radial perturbations of a viscous Newtonian fluid according to Navier-Stokes theory.

The Israel-Stewart case is depicted in Fig.~\ref{fig:IS_Newtonian_shear_bulk}. The presence of the relaxation time $\tau_\pi$ in Eq.~\eqref{eq:wcr_Newt_eta} restricts the critical frequency to the range $\omega_{\textrm{cr}} \in (\omega_\pi,0)$. 
In this case, there are three families of modes for each overtone number $n$: two weakly damped, oscillatory modes (referred to as ``hydrodynamic modes''), and one strongly damped, non-oscillatory mode (referred to as ``non-hydrodynamic''\footnote{A hydrodynamic mode is typically defined as one for which $\lim_{k \to 0} \omega(k) = 0$, whereas for nonhydrodynamic modes $\lim_{k \to 0} \omega(k) \neq 0$. Although our analysis does not involve a continuous dispersion relation, we adopt this standard terminology and classify modes based on their behavior at low $n$.}). In particular, the frequency of the $n=0$ non-hydrodynamic mode is precisely $\omega_\pi$. 
As the overtone number increases, the (imaginary) frequency of non-hydrodynamic modes approaches $\omega_{\textrm{cr}}$, while the imaginary part of the hydrodynamic mode frequencies asymptotes to a larger constant value. 

As in the Navier-Stokes case, the qualitative behavior of $\omega_{(n)}$ in terms of the overtone number $n$ for Israel-Stewart mirrors that of a homogeneous fluid, where $\omega(k)$ are roots of the cubic polynomial equation~\eqref{eq:homogeneous_IS_shear}. To facilitate comparison with our discrete spectrum, we adopt the following ansatz, inspired by the structure of Eq.~\eqref{eq:homogeneous_IS_shear}:
\begin{equation} \label{eq:omega_homogeneous_ansatz_IS_S}
    [\omega^2-\omega_\textrm{PF}^2(n)](1-\omega/\omega_\pi) + 2i\omega \,\omega_\textrm{PF}(n) \left(\frac{n}{n_\textrm{visc}}\right) = 0.
\end{equation}
Since the definition of $k_\textrm{visc}$ in Eq.~\eqref{eq:kvisc} is independent of $\tau_\pi$, we adopt the same prescription for $n_\textrm{visc}$ as before: it is the threshold overtone number  above which modes in the $\tau_\pi = 0$ case become purely damped. As shown in Fig.~\ref{fig:IS_Newtonian_shear_bulk}, the numerical results for the mode frequencies $\omega_{(n)}$ (blue circles) closely follow the roots of the polynomial equation \eqref{eq:omega_homogeneous_ansatz_IS_S} (shaded region).

\subsubsection{Bulk viscosity only}

We now turn to the case of bulk viscosity only, for which the frequency-domain master equation for the Lagrangian displacement, Eq.~\eqref{eq:xiQ0newtonian}, becomes
\begin{align} \label{eq:newtonian_master_bulk}
    0 & = \xi''+ 2 \cot\bar{r} \xi' + \frac{\xi}{2\bar{r}^2 (\bar{\omega} - \bar{\omega}_{\textrm{cr}}) \bar{\omega}_\Pi} \left[ 8 \bar{\omega} \bar{\omega}_{\textrm{cr}} + 4 \bar{\omega}_\Pi \bar{\omega}_{\textrm{cr}} \right.
    \nonumber \\ 
    &  - 12 \bar{\omega} \bar{\omega}_\Pi +     
    \bar{r}^3 \bar{\omega}^2 (\bar{\omega} - \bar{\omega}_\Pi) \bar{\omega}_{\textrm{cr}} \csc \bar{r} 
    \nonumber \\
    & 
    + \left. 8 \bar{r} \bar{\omega} (\bar{\omega}_\Pi - \bar{\omega}_{\textrm{cr}}) \cot \bar{r}  
    \right] ,
\end{align}
where we defined,
\begin{equation}\label{eq:wcr_Newt_zeta}
    \omega_{\textrm{cr}} = -\frac{i \Gamma}{\tau_\zeta + \Gamma \tau_\Pi}, \qquad \omega_\Pi \equiv - \frac{i}{\tau_\Pi}.
\end{equation}

For $\omega \neq \omega_{\textrm{cr}}$, Eq.~\eqref{eq:newtonian_master_bulk} is regular in the interior of the integration domain, $\bar{r} \in (0, \pi)$, and, at the boundaries, regularity of physical solutions demands that $\xi(0) = 0$ and
\begin{equation}
    \xi'(\pi) = \frac{ \bar{\omega} \xi(\pi) \left[ 8 (\bar{\omega}_{\textrm{cr}} - \bar{\omega}_\Pi) + \pi^2 \bar{\omega} \bar{\omega}_{\textrm{cr}} (\bar{\omega} - \bar{\omega}_\Pi)
    \right]}{4\pi \bar{\omega}_\Pi (\bar{\omega} - \bar{\omega}_{\textrm{cr}})}.
\end{equation}

For Navier-Stokes theory (i.e., $\tau_\Pi = 0$), perturbations of a homogeneous fluid lead to the same dispersion relation, Eq.~\eqref{eq:NShomogeneous}, for both bulk and shear viscosity. In the case of radial perturbations of a Newtonian star, the equations governing bulk and shear viscosity cannot be reduced to one another by a simple rescaling of the transport coefficients, as in the homogeneous case. Nevertheless, the ansatz \eqref{eq:omega_homogeneous_ansatz} continues to capture the main qualitative features of the spectrum in the bulk viscous case, as can be seen in Fig.~\ref{fig:NS_Newtonian_shear_bulk} (red triangles).

\begin{figure}[th] 
    \includegraphics[width=\linewidth]{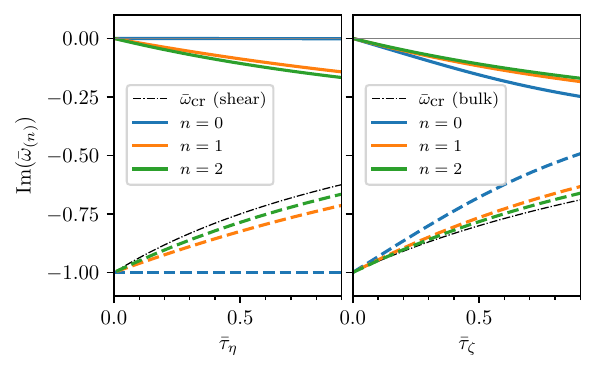}
    \caption{Imaginary part of the eigenfrequencies $\bar{\omega}_n = \omega_{(n)}/\omega_0$ for a Newtonian $\Gamma = 2$ polytrope subject to shear (left panel) or bulk (right panel) viscosity in Israel-Stewart theory, shown as a function of $\bar{\tau}_\eta = \omega_0 \tau_\eta$ or $\bar{\tau}_\zeta = \omega_0 \tau_\zeta$. For the shear-viscous case (left), we set $\omega_\pi = -i\omega_0$, and recall that $\omega_\textrm{cr}$ is defined in Eq.~\eqref{eq:wcr_Newt_eta}. In the bulk viscous case (right panel), we similarly set $\omega_\Pi = -i\omega_0$, with $\omega_\textrm{cr}$ defined in Eq.~\eqref{eq:wcr_Newt_zeta}. Branches of hydrodynamic and non-hydrodynamic modes are shown with solid and dashed lines, respectively. While shear viscosity has a minimal impact on the fundamental ($n=0$) radial modes, bulk viscosity strongly modifies both the hydrodynamic and non-hydrodynamic $n=0$ branches.}
    \label{fig:IS_var_tau}
\end{figure}

The situation changes significantly in the context of Israel-Stewart theory. Although the dispersion relations for bulk and shear viscosity in the case of a homogeneous fluid are similar within this framework  [cf.~Eqs.~\eqref{eq:homogeneous_IS_shear} and \eqref{eq:homogeneous_IS_bulk}], the simple extension \eqref{eq:omega_homogeneous_ansatz_IS_S} no longer captures the qualitative behavior of radial perturbations subject to bulk viscosity in a Newtonian star, particularly at low overtone numbers, as seen in Fig.~\ref{fig:IS_Newtonian_shear_bulk}. The dependence of $\textrm{Im}(\omega_{(n)})$ on $\tau_\eta$ and $\tau_\zeta$ is shown in Fig.~\ref{fig:IS_var_tau} for the lowest values of $n$, where the distinct effects of shear and bulk viscosity on the fundamental ($n=0$) modes can be appreciated.

It is indeed remarkable that the fundamental ($n=0$) modes of the shear-viscous case are so well approximated by the infinite-wavelength ($k=0$) limit of Eq.~\eqref{eq:homogeneous_IS_shear}, despite the expectation that the star's finite size should restrict the maximum wavelength of perturbations. In contrast, the bulk-viscous case exhibits a much stronger sensitivity to finite-size effects, and the identification $(k/k_\textrm{visc}) \to (n/n_\textrm{visc})$ used in Eq.~\eqref{eq:omega_homogeneous_ansatz_IS_S}, which mainly affects the imaginary part of the mode frequencies, is no longer appropriate for bulk viscosity when $n$ is small. 

\begin{figure*}[th] 
    \begin{subfigure}[b]{0.5\textwidth}
        \centering
        \includegraphics[width=\linewidth]{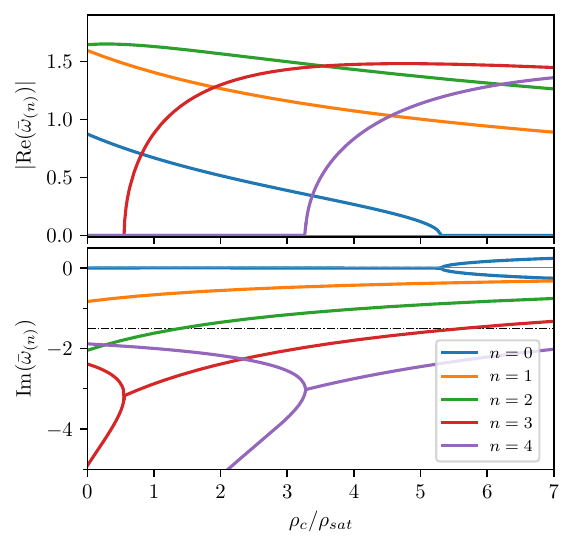}
        \caption{Shear (Navier-Stokes)}
        \label{fig:NS_shear_R}
    \end{subfigure}%
    ~ 
    \begin{subfigure}[b]{0.5\textwidth}
        \centering
        \includegraphics[width=\linewidth]{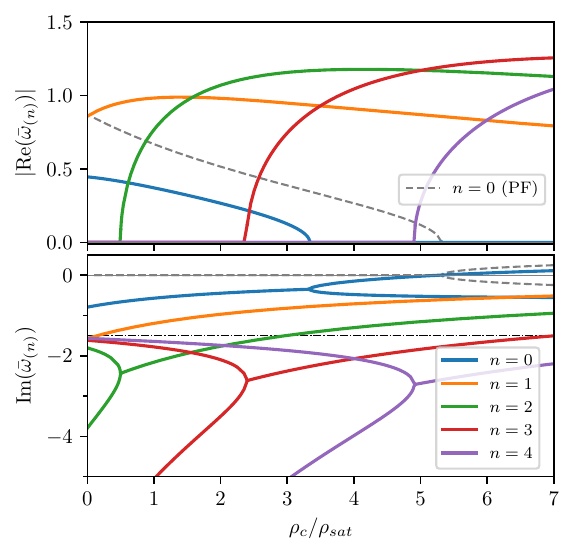}
        \caption{Bulk (Navier-Stokes)}
        \label{fig:NS_bulk_R}
    \end{subfigure}
    \caption{Real (upper panels) and imaginary (lower panels) parts of the eigenfrequencies $\bar{\omega}_{(n)} = \omega_{(n)}/\omega_0$, with $\omega_0 = \sqrt{2\pi\rho_c}$, for $n \in\{0,1,2,3,4\}$, as a function of the central rest-mass density $\rho_c$ (in units of $\rho_\textrm{sat} = 2.7 \times 10^{14} \textrm{g}/\textrm{cm}^3$), for a relativistic $\Gamma = 2$ polytrope within Navier-Stokes theory. Results for shear and bulk viscosity are presented in the left and right panels, respectively, for (a) $\bar{t}_\eta = t_\eta \omega_0 = 1$ and (b) $\bar{t}_\zeta = t_\zeta \omega_0 = 4/3$. A black dot-dashed line in the lower panels indicates the imaginary part of $\bar{\omega}_\textrm{cr} = \omega_\textrm{cr}/\omega_0 = - 1.5 i$, with $\omega_\textrm{cr}$ defined in Eq.~\eqref{eq:wcr_rel}.} 
    \label{fig:NS_R}
\end{figure*}

\subsection{Relativistic analysis} \label{sec:relativistic_results}

Having developed intuition through the analysis of a homogeneous fluid and a Newtonian star, we now turn to the relativistic case. A key difference expected in this regime is a nontrivial dependence on the central density -- or, equivalently, on the star’s compactness, which governs the strength of relativistic effects. In contrast, the Newtonian solutions (for a $\Gamma = 2$ polytrope) discussed earlier are self-similar: both the background and the perturbations can be expressed in terms of dimensionless quantities that satisfy equations independent of the central density. 
For example, the eigenfrequencies for a given $\rho_c$ follow a simple scaling relation, $\omega_{(n)} = \bar{\omega}_{(n)} \omega_0$, where $\omega_0 = \sqrt{2\pi \rho_c}$ is a density-dependent factor and $\bar{\omega}_{(n)}$ are density-independent (dimensionless) eigenfrequencies computed in Sec.~\ref{sec:Newtonian_results}. 
Moreover, our results assumed a fixed value of $\bar{\tau}_\eta =\tau_{\eta} \omega_0$ (or $\bar{\tau}_\zeta = \tau_{\zeta} \omega_0$). If, on the other hand, microscopic considerations fix the timescales $\tau_{\eta, \zeta}$, our Newtonian analysis already suggests that increasing central density effectively leads to a larger viscosity (since it amounts to a larger value of $\bar{\tau}_{\eta,\zeta}$).

\begin{figure*}[th] 
    \begin{subfigure}[b]{0.5\textwidth}
        \centering
        \includegraphics[width=\linewidth]{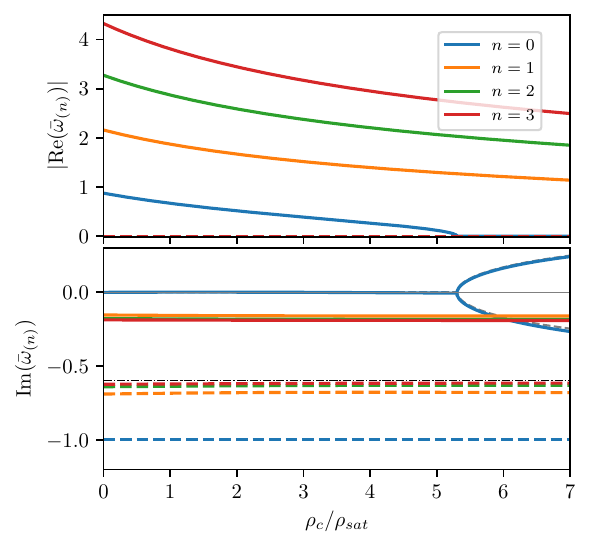}
        \caption{Shear (Israel-Stewart)}
        \label{fig:IS_shear_R}
    \end{subfigure}%
    ~ 
    \begin{subfigure}[b]{0.5\textwidth}
        \centering
        \includegraphics[width=\linewidth]{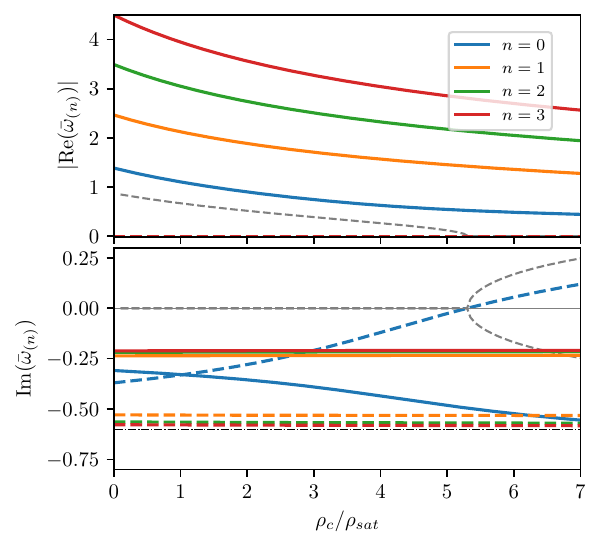}
        \caption{Bulk (Israel-Stewart)}
        \label{fig:IS_bulk_R}
    \end{subfigure}
    \caption{Real (upper panels) and imaginary (lower panels) parts of the eigenfrequencies $\bar{\omega}_{(n)} = \omega_{(n)}/\omega_0$, with $\omega_0 = \sqrt{2\pi\rho_c}$, for $n \in\{0,1,2,3\}$, as a function of the central rest-mass density $\rho_c$ (in units of $\rho_\textrm{sat} = 2.7 \times 10^{14} \textrm{g}/\textrm{cm}^3$), for a relativistic $\Gamma = 2$ polytrope within Israel-Stewart theory. Results for shear and bulk viscosity are presented in the left and right panels, respectively, for (a) $t_\pi \omega_0 = 1$ and $t_\eta \omega_0 = 1$, and (b) $t_\Pi \omega_0 = 1$ and $t_\zeta \omega_0 = 4/3$. These values are such that condition \eqref{eq:lin-causality} is satisfied for the entire range of central densities shown in the plot. Branches of hydrodynamic and non-hydrodynamic modes are shown with solid and dashed lines, respectively. A black dot-dashed line in the lower panels indicates the imaginary part of $\bar{\omega}_\textrm{cr} = \omega_\textrm{cr}/\omega_0 = - 0.6 i$ [cf.~Eq.~\eqref{eq:wcr_rel}]. The fundamental ($n=0$) mode frequencies of a perfect fluid are represented as dashed gray lines for comparison.} 
    \label{fig:IS_R}
\end{figure*}

In the relativistic case, the self-similarity present in the Newtonian limit is lost, and a more intricate dependence on the central density could be expected.

\subsubsection{A minimal (singularity-avoiding) extension} \label{sec:minimal_extension}

We begin our relativistic analysis by assuming a minimal extension of the setup considered in Sec.~\ref{sec:Newtonian_results}. Namely, we consider a relativistic $\Gamma = 2$ polytrope, with the pressure given by Eq.~\eqref{eq:polytropic}, with $K = 0.06 c^2/\rho_\textrm{sat}$, where $\rho_\textrm{sat} = 2.7 \times 10^{14} \textrm{g}/\textrm{cm}^3$ (which yields a $1.92M_\odot$ maximum mass), and the energy density given by
\begin{equation}
    \textrm{e} = \rho + \frac{\textrm{p}}{\Gamma - 1}.
\end{equation}
Moreover, we adopt a similar prescription for the transport coefficients, namely 
\begin{equation} \label{eq:prescription_eta_zeta}
  \eta = \tau_\eta \textrm{p}, \qquad \zeta = \tau_\zeta \textrm{p}.
\end{equation}
However, instead of treating $\tau_{i} \in \{\tau_\pi, \tau_\eta, \tau_\Pi, \tau_\zeta\}$ as constants, we parametrize them as
\begin{equation} \label{eq:tdef}
    \tau_i = t_i e^{\Phi/2},
\end{equation}
with the redshifted relaxation times $t_i \in \{t_\pi, t_\eta, t_\Pi, t_\zeta\}$ assumed to be constant instead. 
In the Newtonian limit, $t_i = \tau_i$ and we recover the rule employed in Sec.~\ref{sec:Newtonian_results}. 

The above prescription for the transport coefficients, although not based on microscopic considerations, greatly simplifies the numerical integration of frequency-domain equations by ensuring that no singular points arise within the interior of the integration domain. In particular, the principal part of (the frequency-domain version of) Eq.~\eqref{eq:xiQ0} vanishes only at the critical frequency $\omega_\textrm{cr}$ satisfying
\begin{equation} \label{eq:wcr_rel}
    \Gamma + \frac{4 \omega_\textrm{cr} t_\eta}{3(\omega_\textrm{cr} t_\pi + i)} + \frac{\omega_\textrm{cr} t_\zeta}{\omega_\textrm{cr} t_\Pi + i} = 0,
\end{equation}
which has the exact same structure as its Newtonian counterpart, Eq.~\eqref{eq:wcr_Newt}, but is given in terms of the redshifted relaxation times $t_i$.

Figure \ref{fig:NS_R} shows the eigenfrequencies ${\omega}_{(n)}$ for the first overtone numbers $n$ as a function of central density, within Navier-Stokes theory. To facilitate comparison with the Newtonian case, we assume fixed values of the rescaled timescales $\bar{t}_{\eta} = t_{\eta}\omega_0$ and $\bar{t}_{\zeta} = t_{\zeta}\omega_0$, where $\omega_0 = \sqrt{2\pi\rho_c}$ depends on the central density. In the Newtonian limit, such a choice would guarantee that the rescaled eigenfrequencies $\bar{\omega}_{(n)} = {\omega}_{(n)}/\omega_0$ are independent of $\rho_c$. Therefore, any remaining density dependence observed in Fig.~\ref{fig:NS_R} can be attributed to relativistic effects.

For the value of $\bar{t}_{\eta}$ used in the left panel of Fig.~\ref{fig:NS_R}, only the $n=0,1,2$ modes are oscillatory in the Newtonian ($\rho_c \to 0$) limit, corresponding to $n_\textrm{visc} \lesssim 3$ in Eq.~\eqref{eq:omega_homogeneous_ansatz}. Higher overtones appear as pairs of purely damped modes. In the relativistic setting, as the central density increases, these purely damped modes can transition back into oscillatory behavior. The higher the overtone number, the larger the central density required for this transition to occur, in such a way that $n_\textrm{visc}$ increases with central density\footnote{We emphasize that this holds for a fixed $\bar{t}_\eta$. If ${t}_\eta$ was held fixed instead, $n_\textrm{visc}$ would instead decrease with increasing $\rho_c$.}. The basic qualitative view provided by Eq.~\eqref{eq:omega_homogeneous_ansatz} -- with $n_\textrm{visc}$ depending on the central density -- remains valid in the relativistic setting until close to the onset of radial instability. As $\rho_c$ approaches the critical value for radial instability (in our case, $\rho_{c,\textrm{cr}} \approx 5.3 \rho_\textrm{sat}$), the frequency of the $n=0$ modes become purely imaginary, with one branch becoming unstable above $\rho_{c,\textrm{cr}}$.

The case of bulk viscosity, shown in Fig.~\ref{fig:NS_bulk_R}, is qualitatively similar to the shear-viscous case, but with a noticeable distinction regarding the behavior of the fundamental ($n=0$) mode. While shear viscosity leaves the fundamental mode frequency practically unaltered, bulk viscosity has a much stronger impact on both its real and imaginary parts. It drives the real part of the $n=0$ mode frequencies to zero, rendering these modes overdamped (purely imaginary frequency) well before the onset of radial instability. As in the shear-viscous case, one of $n=0$ branches becomes unstable above the critical central density $\rho_{c,\textrm{cr}}$. Viscosity does not shift the critical central density itself -- and therefore whether the star is stable or not \cite{Caballero:2025omv} -- but naturally reduces the instability timescale -- much more effectively in the bulk-viscous case. 

We now turn to the Israel-Stewart theory, with results for a relativistic $\Gamma = 2$ polytrope presented in Fig.~\ref{fig:IS_R}. In the shear-viscous case (Fig.~\ref{fig:IS_shear_R}), the (rescaled) imaginary part of the eigenfrequencies $\bar{\omega}_{(n)}$ shows a remarkable independence from the central density. In particular, the fundamental mode frequency is only slightly affected by shear viscosity, similarly to what is observed in the Navier-Stokes case. In contrast, bulk viscosity strongly affects the fundamental mode frequency, as shown in Fig.~\ref{fig:IS_bulk_R}: the real part of the $n=0$ hydrodynamic mode frequencies no longer vanishes near $\rho_{c,\textrm{cr}}$ and, instead, it is the $n=0$ non-hydrodynamic mode that becomes unstable for $\rho_c > \rho_{c,\textrm{cr}}$. 

\begin{figure}[th] 
    \includegraphics[width=\linewidth]{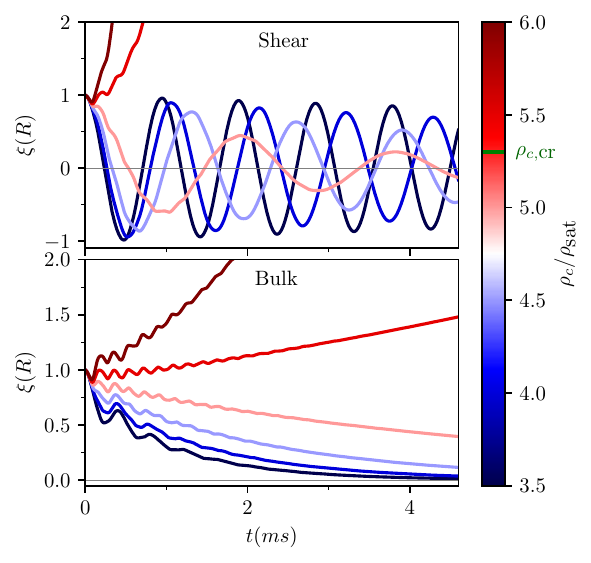}
    \caption{Lagrangian displacement evaluated at the stellar surface for perturbations including shear (upper panel) or bulk (lower panel) viscosity, as a function of time. Different colors correspond to different background central density configurations. We adopt a $\Gamma = 2$ polytropic equation of state, and set $t_{\eta} = t_{\pi} = 0.2$ ms in the upper panel and $t_{\zeta} = t_{\Pi} = 0.2$ ms in the bottom panel. }
    \label{fig:time_evol}
\end{figure}

To further highlight the contrasting roles of bulk and shear viscosity, we display in Fig.~\ref{fig:time_evol} the time evolution of the Lagrangian displacement at the stellar surface, $\xi(t,r=R)$, computed within Israel-Stewart theory. These results are obtained by numerically integrating Eqs.~\eqref{eq:xiQ0} and \eqref{eq:IS-eoms-Q0} with the following initial conditions:
\begin{equation}
    \xi(t=0,r) = r/R, \qquad \dot{\xi}(t=0,r) = 0,
\end{equation}
and assuming $\pi(t=0,r) = 0$ and $\Pi(t=0,r) = 0$. 
Under these initial data, the dynamics is expected to be dominated by the fundamental mode.

Figure~\ref{fig:time_evol} shows that, in the shear-viscous case, oscillations are only weakly damped, consistent with the small values of $\operatorname{Im}(\omega)$ predicted for the $n=0$ hydrodynamic mode. As the critical density $\rho_{c,\textrm{cr}}$ is approached, the oscillation frequency decreases, and the fundamental mode becomes unstable beyond this threshold. Bulk viscosity, in contrast, leads to a qualitatively different behavior. For the chosen values of $\tau_\zeta$ and $\tau_\Pi$, the mode structure resembles that in Fig.~\ref{fig:IS_bulk_R}: the longest-lived mode is the nonhydrodynamic $n=0$ mode, which drives a purely exponential decay of perturbations that transitions to exponential growth as $\rho_{c,\textrm{cr}}$ is crossed. Fits to the time series confirm the dominance of this nonoscillatory mode, with additional contributions from the oscillatory hydrodynamic $n=0$ and $n=1$ modes.

\subsubsection{(Singularity-prone) Generalizations}

We conclude this section by revisiting some of the simplifying assumptions previously made regarding the choice of EoS and transport coefficients, while leaving further generalizations to future work.

\begin{figure}[th] 
    \includegraphics[width=\linewidth]{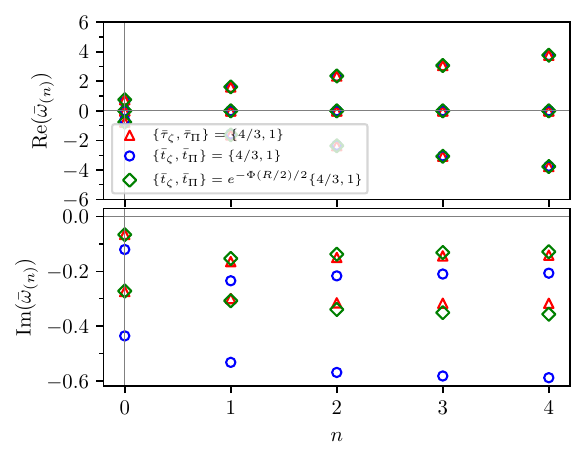}
    \caption{Real (upper panel) and imaginary (lower panel) parts of the eigenfrequencies $\bar{\omega}_{(n)} = \omega_{(n)}/\omega_0$ as a function of the overtone number $n$ for a relativistic $\Gamma = 2$ polytropic star with central density $\rho_c = 4 \rho_\textrm{sat}$, subject to bulk viscosity within the Israel-Stewart theory. Red triangles correspond to the constant-$\tau_i$ prescription, with $\bar{\tau}_\zeta = \tau_\zeta \omega_0 = 4/3$ and $\bar{\tau}_\Pi = \tau_\Pi \omega_0 = 1$. Blue circles and green diamonds correspond to the constant-$t_i$ prescription, where $t_i = \tau_i$ (i.e., $\bar{t}_\zeta = 4/3$ and $\bar{t}_\Pi = 1$) and $t_i = e^{-\Phi(R/2)/2} \tau_i$ (i.e., $\bar{t}_\zeta \approx 2.17$ and $\bar{t}_\Pi = 1.63$), respectively. Accounting for a mean redshift factor results in good overall agreement between the two prescriptions.}
    \label{fig:IS_comparison}
\end{figure}

We begin by contrasting the prescription in which the redshifted relaxation times $t_i$ are held constant with that in which the proper relaxation times $\tau_i$ are fixed instead. 
The constant-$t_i$ prescription lacks a firm physical justification, as it implies that the proper timescales $\tau_i$ would depend on the gravitational potential---and hence on global spacetime properties---whereas they are expected to depend only on local thermodynamical quantities. Still, since the redshift factor $e^{\Phi(r)/2}$ is an order-of-one quantity for neutron stars, both prescriptions yield qualitatively similar results, as illustrated in Fig.~\ref{fig:IS_comparison} for the case of bulk viscosity within the Israel-Stewart theory. A better quantitative agreement between the two approaches can be achieved by introducing a mean redshift factor---such as $e^{-\Phi(R/2)/2}$---when mapping parameters from one prescription to the other. This effective time dilation of microscopic timescales is a conceptually appealing relativistic correction to the Newtonian picture discussed earlier. 

\begin{figure}[th] 
    \includegraphics[width=\linewidth]{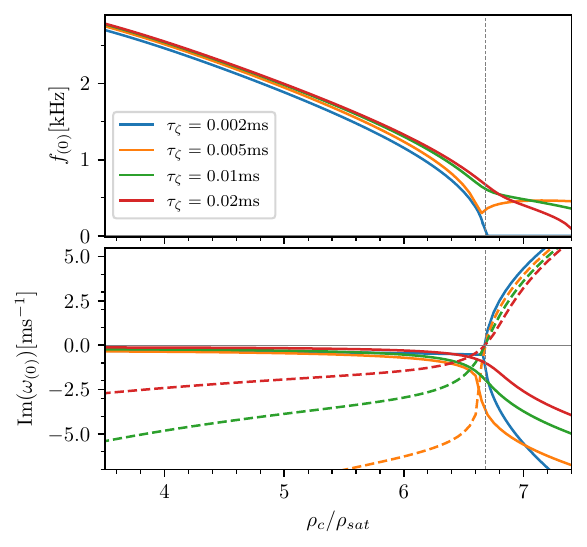}
    \caption{Frequency $f_{(0)} = |\operatorname{Re}(\omega_{(0)})|/(2\pi)$ (upper panel) and inverse damping time $\operatorname{Im}(\omega_{(0)})$ (lower panel) of the $n=0$ radial modes as functions of central density, in the presence of bulk viscosity in the Israel-Stewart theory, using the SLy9 EoS. Branches of hydrodynamic and non-hydrodynamic modes are shown with solid and dashed lines, respectively. In all cases, we set $\tau_\Pi = 10 \tau_\zeta$. The nonhydrodynamic mode is responsible for the onset of instability to gravitational collapse for sufficiently large values of $\tau_{\zeta,\Pi}$.}
    \label{fig:IS_SLY9}
\end{figure}

Next, in Fig.~\ref{fig:IS_SLY9}, we present results for a more realistic EoS, employing a piecewise polytropic approximation to the SLy9 EoS based on the parametrization introduced in Ref.~\cite{OBoyle:2020qvf}. Using a realistic---typically stiffer---EoS alters the speed of sound and, consequently, the timescales required to ensure causal behavior. In this work, we adopt Eq.~\eqref{eq:lin-causality} as a guiding criterion for causality within Israel-Stewart theory, though it is not derived self-consistently from our perturbation equations. This expression imposes a density-dependent constraint on the combination of transport coefficients $4 \tau_\eta / (3 \tau_\pi) + \tau_\zeta / \tau_\Pi$. In Fig.~\ref{fig:IS_SLY9}, we focus on the case of bulk viscosity alone, fixing the ratio $\tau_\zeta/\tau_\Pi = 0.1$, which ensures that Eq.~\eqref{eq:lin-causality} is satisfied throughout the relevant range of densities. 
As for a polytropic EoS, we see that in the bulk-viscous case it is often the nonhydrodynamic mode that becomes unstable at the critical central density corresponding to the maximum mass configuration of the equilibrium sequence.

\section{Conclusions}
\label{sec:conclusion}

We have investigated the behavior of radial oscillations of relativistic viscous stars. Several formulations of relativistic dissipative hydrodynamics are available; in this work, we employed the Navier-Stokes and Israel-Stewart theories, leaving alternative approaches, such as the BDNK theory, for future investigations.
After discussing the gravity-fluid equations of motion in the nonlinear regime, we derived the linearized equations governing radial perturbations. In the linear regime, and assuming vanishing energy diffusion, we obtained a master equation for the Lagrangian displacement [Eq.~\eqref{eq:xiQ0}], coupled to the Israel-Stewart equations for shear and bulk viscosity [Eq.~\eqref{eq:IS-eoms-Q0}]. 

Assuming a harmonic time dependence for the perturbation variables, the problem reduces to an eigenvalue problem, from which the oscillation spectrum $\{\omega_{(n)}\}$ can be determined. Interestingly, we found that poles may appear in the coefficients of the frequency-domain master equation when the mode frequency is purely imaginary; a feature that seems to be generic. In the presence of such singularities, solutions can still be obtained; however, they are discontinuous and have support only on one side of the singular point. Since the perturbation equations would be modified in the presence of energy diffusion, it would be interesting to investigate whether this alters their singularity structure -- a possibility we leave for future work. 

Insight into the mode spectrum can be gained from the dispersion relations of perturbations in an infinite, homogeneous fluid in flat spacetime. In Navier-Stokes theory, there are only two families of hydrodynamic modes, which transition from oscillatory to purely damped behavior at a critical wavelength. In contrast, the Israel-Stewart framework, where shear and bulk viscosity are treated as independent dynamical degrees of freedom, admits additional families of nonhydrodynamic modes, characterized by a dispersion relation such that $\lim_{k\to 0}\omega(k) \neq 0$. By first examining Newtonian, and then relativistic stars, we found that much of that behavior carries over to the stellar context, but with a discretized dispersion relation imposed by the system's finite size [see Fig.~\ref{fig:Newtonian}]. 

For $n\geq 1$ overtones, the effects of bulk and shear viscosity are of comparable magnitude. However, shear viscosity is shown to leave the fundamental ($n=0$) modes largely unchanged, whereas bulk viscosity strongly affects them (cf.~Fig.~\ref{fig:IS_var_tau}) -- a behavior consistent with the symmetry of the problem. The influence of bulk viscosity on the $n=0$ modes is also shown to be highly sensitive to relativistic corrections. This is illustrated for Navier-Stokes theory in Fig.~\ref{fig:NS_bulk_R} and for Israel-Stewart theory in Figs.~\ref{fig:IS_bulk_R} and \ref{fig:IS_SLY9}, the latter employing a more realistic EOS. In Israel-Stewart theory, we further show that, depending on the strength of bulk-viscous effects, stability against gravitational collapse may be determined by the $n=0$ nonhydrodynamic mode rather than the hydrodynamic one. The strikingly different impacts of shear and bulk viscosity can also be visualized in the time-domain simulation shown in Fig.~\ref{fig:time_evol}. 
As far as unstable radial modes are concerned, we find that viscosity acts to soften the instability (i.e., it increases the growth timescale of the perturbations); however, it is not able to suppress it altogether, consistent with the conclusion from the energy-based stability analysis in Ref.~\cite{Caballero:2025omv}.

In this work, we adopted a simplified yet physically reasonable prescription for the transport coefficients. In particular,  we assumed that the shear and bulk viscosities, $\eta$ and $\zeta$, are proportional to the thermodynamic pressure [see Eq.~\eqref{eq:eta-zeta-param}]. Two prescriptions were considered for the relaxation times $\tau_{\pi}$ and $\tau_{\Pi}$, as well as for the viscosity-related timescales $\tau_{\eta}$ and $\tau_{\zeta}$: one in which these quantities are held constant, and another in which the redshifted timescales $t_i = \tau_i e^{-\Phi/2}$ are fixed. The latter approach, combined with a polytropic equation of state, has the technical advantage of avoiding singular points in the integration domain. The two prescriptions can be approximately related by introducing a mean redshift factor into the mapping between them, as illustrated in Fig.~\ref{fig:IS_comparison}.

While the precise form of the equation of state and transport coefficients is crucial for quantitative modeling—and transport coefficients, being tied to non-equilibrium microphysics, represent a major source of uncertainty—the qualitative picture developed here provides a robust physical interpretation of the oscillation spectrum of viscous stars. We expect these insights to remain valid beyond the simplified setting considered, serving as a foundation for understanding not only radial oscillations but also nonradial modes, which are more directly tied to gravitational-wave emission.

\section*{Acknowledgments}

R.M.~acknowledges financial support from CNPq (Conselho Nacional de Desenvolvimento Científico e Tecnológico) and FAPERJ (Fundação Carlos Chagas Filho de Amparo à Pesquisa do Estado do Rio de Janeiro), Grant E-26/204.589/2024. A.~G. and J.~M.~ acknowledge financial support from CAPES (Coordenação de Aperfeiçoamento de Pessoal de Nível Superior). G.~S.~R. is partially supported by Vanderbilt University and by the U.S. Department of Energy, Office of Science under Award Number DE-SC-0024347. G.S.D.~acknowledges CNPq as well as FAPERJ, Grant No.~E-26/202.747/2018.

\appendix

\section{Background structure equations} \label{sec:background}

The hydrostatic equilibrium equations for a relativistic star are given by
\begin{align}
    \Lambda' & = \frac{e^\Lambda}{r} (e^{-\Lambda}-1 + 8\pi r^2 \textrm{e}), \\
    \Phi' & = \frac{e^\Lambda}{r} (1-e^{-\Lambda} + 8 \pi r^2 \textrm{p}), \\
    \textrm{p}' & = (\textrm{e} + \textrm{p}) \frac{\Phi'}{2},
\end{align}
and can be recast in terms of the mass aspect function $m = (r/2)(1-e^{-\Lambda})$. Given a one-parameter equation of state, $\textrm{p}=\textrm{p}(\textrm{e})$, these equations can be numerically integrated outward from $r=0$, where $\Lambda(0) = 0$, up to the stellar surface $r = R$, defined by $\textrm{p}(R) = 0$. Continuity with the exterior Schwarzschild solution then implies $e^{\Phi(R)} = 1 - 2M/R$, where $M = m(R)$ is the total mass. These boundary conditions still allow freedom in the choice of  $p_c = \textrm{p}(0)$, such that varying $p_c$ generates a one-parameter sequence of equilibrium configurations.

In the Newtonian limit, the structure equations become
\begin{equation} \label{eq:newtonian-structure-eqs}
    \textrm{p}' = - \frac{m\rho}{r^2}, \qquad
    m' = 4\pi r^2 \rho,
\end{equation}
where $\rho$ denotes the rest-mass density.
Analytic solutions to the Newtonian structure equations can be obtained in certain cases, such as for a $\Gamma = 2$ polytropic equation of state of the form \eqref{eq:polytropic}. In this case, we obtain
\begin{equation}
    \textrm{p}(\bar{r}) = p_c \left(\frac{\sin \bar{r}}{\bar{r}} \right)^2, \qquad
    \bar{m}(\bar{r}) = \sin \bar{r} - \bar{r} \cos\bar{r},
\end{equation}
where $\bar{r} \equiv r/r_0$ and $\bar{m} = m/m_0$ are dimensionless variables, with $r_0^2 \equiv p_c/(2\pi \rho_c^2)$ and $m_0 \equiv 4\pi \rho_c r_0^3$. An additional frequency scale will be useful in the frequency domain analysis presented in the main text, and for that purpose, we define $\omega_0 = \sqrt{2\pi \rho_c}$.
The stellar radius occurs at $\bar{r} = \pi$, at which point pressure vanishes.

\section{Longitudinal perturbations around a homogeneous fluid in a flat spacetime}
\label{apn:long-pert-hom-fluid}

The discussion in this section summarizes what is presented in Ref.~\cite[cap.~2]{Denicol:2021}. We consider a fluid initially in global equilibrium, such that its state is specified by $\{\mathrm{n}_{(0)}, \mathrm{e}_{(0)}, \mathrm{p}_{(0)}(\mathrm{n}_{(0)}, \mathrm{e}_{(0)}), u^{\mu}_{(0)}\}$. Then, it is linearly perturbed to a non-equilibrium state specified by $(\mathrm{n}_{(0)} + \bar\delta \mathrm{n}, \mathrm{e}_{(0)} + \bar\delta \mathrm{e} , \mathrm{p}_{(0)}(\mathrm{n}_{(0)}, \mathrm{e}_{(0)}) + \bar\delta \mathrm{p}_{(0)} + \bar\Pi, u^{\mu}_{(0)} + \bar\delta u^{\mu}, \bar{\pi}^{\alpha \beta})$. Then, from the linearized local conservation laws \eqref{eq:basic-hydro-EoM} in Fourier space, in a regime where 
energy diffusion perturbations vanish, we have
\begin{subequations}
\label{eq:hydro-EoMs-landau-Fourier}
\begin{align}
\label{eq:hydro-EoM-eps-landau-Fourier}
 - i \Omega \bar\delta \mathrm{e}  + (\mathrm{e}_{(0)}+ \mathrm{p}_{(0)}) i (\kappa_{\alpha} \bar\delta u^{\alpha})  = 0 + \mathcal{O}(\bar\delta^{2}), \\
\label{eq:hydro-EoM-umu-landau-Fourier}
 -\left(\mathrm{e}_{(0)} + \mathrm{p}_{(0)}\right)i \Omega \bar\delta u^{\mu} 
+
i\kappa^{\mu}\left(\bar\delta \mathrm{p} + \bar \Pi \right)  
+ 
i \kappa_{\alpha} \bar \pi^{\alpha \mu}
\notag
\\
= 0 + \mathcal{O}(\bar\delta^{2}),
\end{align}    
\end{subequations}
where $\Omega = - u^{\mu}_{(0)}k_{\mu}$  and $\kappa^{\mu} = (g^{\mu \nu} + u^{\mu}_{(0)} u^{\nu}_{(0)}) k_{\nu}$. In the local rest frame of the background fluid $\Omega$ reduces to the frequency $k^{0} = \omega$ and $\kappa^{\mu}$ reduces to the wave vector, $\kappa^{\mu} = (0, \Vec{k})^{T}$. Once again, the above expressions must be complemented by the constitutive relations/dynamical equations for the stresses, which change according to the dissipative hydrodynamic theory employed. It is convenient to decompose the vector and tensor perturbations $\bar\delta u^{\mu}$ and $\bar \pi^{\alpha \beta}$ with respect to $\kappa^{\mu}$, 
\begin{equation}
\begin{aligned}
&
\bar\delta u^{\mu} = \bar\delta u_{\parallel} \frac{\kappa^{\mu}}{\kappa} + \bar\delta u^{\mu}_{\perp}, 
\\
&
\bar \pi^{\alpha \beta} = \bar \pi_{\parallel}
\left(
\frac{\kappa^{\mu}\kappa^{\nu}}
{\kappa^{2}}
-
\frac{1}{2}
\Delta^{\alpha \beta}_{\kappa}
\right)
+
\bar \pi^{\alpha \beta}_{\perp},
\end{aligned}    
\end{equation}
where $\Delta^{\alpha \beta}_{\kappa} = g^{\alpha \beta} + u_{(0)}^{\alpha} u_{(0)}^{\beta} - (1/\kappa^{2}) \kappa^{\alpha}\kappa^{\beta}$ is the projection in the linear subspace orthogonal to both $u^{\mu}$ and $\kappa^{\mu}$ and $\kappa^{2} = \kappa^{\mu} \kappa_{\mu}$. In the background rest frame $\kappa^{2} = k^{2}$, the squared magnitude of the wave vector and the frequency. Then, from Eqs.~\eqref{eq:hydro-EoMs-landau-Fourier} one can derive a linear system of algebraic equations that only involve longitudinal perturbations $\bar\delta u_{\parallel}$ and $\bar \pi_{\parallel}$ and another involving only transversal perturbations $\bar\delta u^{\mu}_{\perp}$ and $\bar \pi^{\alpha \beta}_{\perp}$. Since we are interested only in the dispersion relations stemming from longitudinal perturbations, we shall consider $\bar\delta u^{\mu}_{\perp} = 0 = \bar \pi^{\alpha \beta}_{\perp}$. This implies that, in the rest frame of the background fluid $\bar\delta u^{i}_{RF} = \bar\delta u_{\parallel} \hat{k}^{i}$, and $\bar\delta\pi^{i j}_{RF} = (3/2) \bar \pi_{\parallel}  [\hat{k}^{i} \hat{k}^{j} - (1/3) \delta^{ij}]$.    

For Navier-Stokes in Fourier space we have $\bar \Pi = - \zeta i \kappa_{\mu} \bar\delta u^{\mu} $,  $\bar \pi^{\alpha \beta} = - 2 i \eta \Delta^{\alpha \beta \mu \nu}_{(0)}\kappa_{\mu} \bar\delta u_{\nu}$, where $\Delta^{\alpha \beta \mu \nu}_{(0)}$ is the doubly-symmetric, traceless projector that is fully orthogonal with respect to the background four-velocity. Hence, Eqs.~\eqref{eq:hydro-EoMs-landau-Fourier} can be expressed as 
\begin{equation}
\left(\begin{array}{cc}
 - i \Omega  & i \kappa \\
 i \kappa c_{s}^{2} & - i \Omega + 2 c_s\kappa^{2}/k_{\rm visc} 
\end{array}    
\right)
\left(\begin{array}{c}
    \bar\delta \mathrm{e} /(\mathrm{e} + \mathrm{p})  \\
    \bar\delta u_{\parallel}  
\end{array}    
\right)
=
\left(\begin{array}{c}
    0   \\
    0  
\end{array}    
\right).
\end{equation}
Then, from the condition that the determinant of the matrix defining the above linear system vanishes,  we obtain the  polynomial equation  
\begin{equation}
\Omega^{2} + 2 i c_{s} \frac{\kappa^{2}}{k_{\rm visc}} \Omega - \kappa^{2} c_{s}^{2} = 0 ,   
\end{equation}
which in the local rest frame of the background fluid can be solved for the frequency and yields the dispersion relation \eqref{eq:NShomogeneous}.

On the other hand, for Israel-Stewart we have, in Fourier space $(1-i \Omega \tau_\Pi)\bar \Pi = - \zeta i\kappa_{\mu} \bar\delta u^{\mu} $, for the bulk relaxation equation and $(1-i \Omega\tau_\pi)\bar \pi^{\alpha \beta} = - 2 i \eta \Delta^{\alpha \beta \mu \nu}_{(0)} \kappa_{\mu} \bar\delta u_{\nu}$ for the shear relaxation equation. For systems with only bulk dissipative perturbations, Eqs.~\eqref{eq:hydro-EoMs-landau-Fourier} can be expressed as
\begin{equation}
\label{eq:matrix-diss-is-pb}
\begin{aligned}
\left(\begin{array}{ccc}
 - i \Omega  & i \kappa & 0 \\
 i \kappa c_{s}^{2} & - i \Omega  & i \kappa \\
 0 & 2ic_s \kappa/k_{\rm visc}\vert_{\eta = 0}  & (-i \Omega \tau_{\Pi} + 1)
\end{array}    
\right)
\left(\begin{array}{c}
    \bar\delta \mathrm{e} /(\mathrm{e} + \mathrm{p})  \\
    \bar\delta u_{\parallel} \\
    \bar \Pi/(\mathrm{e} + \mathrm{p}) 
\end{array}    
\right)
\\
=
\left(\begin{array}{c}
    0   \\
    0   \\
    0
\end{array}    
\right).
\end{aligned}
\end{equation}
The singularity of the matrix defining this linear system of equations leads to
\begin{equation}
\label{eq:dispersion-pb}
(\Omega^2-c_s^2 \kappa^2)(1-i\Omega \tau_\Pi) + 2i\Omega c_s \kappa \left(\frac{\kappa}{k_\textrm{visc}|_{\eta=0}}\right) = 0,
\end{equation}
which in the local rest frame of the background fluid reduces to Eq.~\eqref{eq:homogeneous_IS_bulk}. Conversely, for systems with only shear dissipative perturbations, Eqs.~\eqref{eq:hydro-EoMs-landau-Fourier} can be expressed as  
\begin{equation}
\begin{aligned}
\left(\begin{array}{ccc}
 - i \Omega  & i \kappa & 0 \\
 i \kappa c_{s}^{2} & - i \Omega  & i \kappa \\
 0 & 2ic_s\kappa/k_{\rm visc}\vert_{\zeta = 0}  & (-i \Omega \tau_{\pi} + 1)
\end{array}    
\right)
\left(\begin{array}{c}
    \bar\delta \mathrm{e} /(\mathrm{e} + \mathrm{p})  \\
    \bar\delta u_{\parallel} \\
    \bar \pi_{\parallel}/(\mathrm{e} + \mathrm{p}) 
\end{array}    
\right)
\\
=
\left(\begin{array}{c}
    0   \\
    0   \\
    0
\end{array}    
\right),
\end{aligned}
\end{equation}
which is just Eq.~\eqref{eq:matrix-diss-is-pb}, with $\tau_{\Pi} \mapsto \tau_{\pi}$, $k_{\rm visc}\vert_{\eta = 0} \mapsto k_{\rm visc}\vert_{\zeta = 0}$ and $\bar \Pi \mapsto \bar \pi_{\parallel}$.  Thus, imposing that the determinant of the above-defined matrix vanishes amounts to Eq.~\eqref{eq:dispersion-pb} with $\tau_{\Pi} \mapsto \tau_{\pi}$, $k_{\rm visc}\vert_{\eta = 0} \mapsto k_{\rm visc}\vert_{\zeta = 0}$,
\begin{equation}
\label{eq:dispersion-ps}
(\Omega^2-c_s^2 \kappa^2)(1-i\Omega \tau_\pi) + 2i\Omega c_s \kappa \left(\frac{\kappa}{k_\textrm{visc}|_{\zeta=0}}\right) = 0,      
\end{equation}
which in the local rest frame of the background fluid leads to Eq.~\eqref{eq:homogeneous_IS_shear}.

\bibliography{refs}
\bibliographystyle{apsrev4-2}

\end{document}